\documentclass[12pt]{elsarticle}
\usepackage{amsfonts,hyperref,textcomp}
\usepackage{caption,subcaption}
\usepackage{graphicx}
\usepackage{blkarray}
\usepackage{array}
\usepackage[section]{placeins}
\usepackage{amssymb}
\usepackage{booktabs}
\usepackage{multirow}
\usepackage{float}
\usepackage{epstopdf}
\usepackage{color}
\usepackage{amsmath}
\usepackage[normalem]{ulem}

\usepackage{enumitem}
\newlist{worddefs}{description}{1}
\setlist[worddefs]{font=\sffamily\bfseries, labelindent=\parindent, leftmargin=6em, style=sameline}

\begin{document}

\title{Opinion dynamics modelling: distinct attraction and repulsion topologies highlight quantitative effects of trolling }

\author[rvt]{Jake Boyce}
\author[rvt]{Matteo Farina}
\author[rvr]{Jody McKerral}
\author[rvr]{Sergiy Shelyag}
\author[rvt]{Mathew Zuparic}
\address[rvt]{Defence Science and Technology Group, Canberra, ACT 2600, Australia}
\address[rvr]{College of Science and Engineering, Flinders University, Tonsley, Adelaide, SA 5042, Australia}

\date{\today} 

\begin{abstract}
We introduce a model of opinion dynamics based on networked non-linear differential equations. The model  combines a linear attraction with a repulsive hyperbolic tangent interaction, labeled \textit{controversialness}. For low controversialness the model displays universal consensus, which is typical of opinion models. As controversialness increases, opinion behaviours such as  \textit{polarisation}, \textit{clustering} and \textit{dissensus} emerge, dependent on the network topology. By placing attractive and repulsive interactions on distinct networks, this model is able to simulate the manipulative effects of trolls by introducing controversy, which may be associated with mis/disinformation, toxic messaging, and encouraging provocative questioning and/or emotional posting. This work offers an analytical and statistical analysis of model results, under a wide variety of topologies and initial conditions, whilst also generalising cluster detection algorithms typically applied to discrete models. 
\end{abstract}

\maketitle


\section{Introduction} \label{intro}

In an attempt to understand the effect of social influence on opinions, the two-step influence model was first hypothesised by Lazarsfeld et al. \cite{Lazarsfeld48}, based on results emerging from sociological studies of voting patterns of the 1940 and 1944 U.S. presidential elections. The authors noted that changes in voting behaviour stemmed largely from personal communication with \textit{opinion leaders}, between sources of information and audiences. These privileged individuals directly accessed, interpreted, and disseminated information. This model of persuasion challenged the (then) understanding that society consisted of mostly-disconnected individuals directly influenced by mass media \cite{Katz55}. 

Further research into multi-step models focused on the role of social networks in the diffusion and impact of information \cite{Menzel55}. For example, studies focusing on decision-making patterns in fashion choices \cite{Katz57}, and adoption and prescription of new medications amongst a cohort of doctors \cite{Coleman59} found that behavioural change  largely depended on word-of-mouth from opinion-leaders to small groups, and then to the wider community. In addition to technical competence, opinion leaders personified certain \textit{values}, which those being influenced wanted to emulate \cite{Marsh54}. Furthermore, Rogers \cite{Rogers95} revealed that opinion leadership was a property that emerged, and faded, dynamically amongst individuals. Theories stemming from these studies have since been tested on a range of applications, including sustainable pesticide application \cite{Looby69, Feder06, Vandenberg07}, and science communication to maximise the spread of factual information \cite{Nesbit09}.

The advent of the internet has changed the way that people interact and consume information, with unprecedented levels of speed and access. As a result, everyone can be an opinion leader \cite{Casalo20}. 
One by-product of the ubiquity of information technology is the emergence of trolling behaviour. Though the act of trolling appears to be timeless, as demonstrated by its presence in ancient Rome where individuals used controversial graffiti to provoke others \cite{Balami24}, the 21st century is shaping to be a trolls' godsend due to a combination of online anonymity, and anti-social behaviours correlating with enjoyment and social approval amongst peers \cite{Soares23}. Trolls can use social media platforms to exploit controversial issues and events and easily reach swathes of people, manipulate discussions and generate controversy at an unprecedented scale to foster division and fragment societies \cite{Jane15, Lee05, Farina25, Yuan25}. 
Examples of online trolling include expressing extreme opinions, asking provocative questions or encouraging emotional posting with the intent of upsetting others. More sophisticated trolling behaviour additionally employs disinformation, as witnessed in the 2012 Assam riots of northeast India where an individual disseminated more than 20,000 messages containing fabricated gruesome images which ignited regional religions tensions \cite{Goolsby13}. Special interest groups such as the anti-vaccine movement \cite{Deer20} often employ toxic messaging in online forums regarding vaccines, and encourage other users to express underlying fears to control the narrative and counter messaging from the scientific community \cite{Kata12,Miyazaki24}. Trolling is also a tool for state-actors pursuing \textit{hybrid-warfare}, being actively employed towards Baltic \cite{Spruds15} and Eastern European countries \cite{Snegovaya15} to promote an \textit{anti-EU} agenda, and ultimately challenge the legitimacy of their sovereignty. 

Thus, the literature clearly demonstrates that trolls operate and thrive on networks via \textit{controversialness} --- consisting of toxic messaging regarding mis/disinformation, encouraging provocative questions, and expressing controversial opinions which empower community members to post emotional and upsetting responses, forming divisions within groups. 
This work intends to further the understanding of trolling by offering a quantitative model to understand its effects in a general social network setting. 
The following seeks to contextualise our work through the lens of previous efforts in quantitative modelling of opinion dynamics. For more extensive reviews on the topic refer to \cite{Castellano09, Peralta22, Baumannphd} and references therein. 

Inspired by the empirical work of Lazarsfeld et al. \cite{Lazarsfeld48, Katz55}, Abelson \cite{Abelson64} considered a linear mathematical model of interacting agents with dynamic opinions (akin to an \textit{n-step} opinion model) in the presence of constant mass-media communication sources. Though universal consensus was the most common outcome, Abelson was able to demonstrate configurations that resulted in unequal opinion distributions. Taylor \cite{Taylor68} generalised Abelson's work by introducing nonlinear interaction terms representing agent stubbornness, as well as weakening attraction if opinions are sufficiently far apart. Guided by experimental findings that the effect of controversy is capped \cite{Jayles17}, and recognising that consensus in large groups is uncommon since the advent of social media, Baumann et al. \cite{Baumann20} introduced the notion  of controversy to a simplified Abelson model that captured the link between echo chambers and polarisation, in the presence of a stochastically dynamic network. The authors extended the model in \cite{Baumann21} to a multidimensional vector of opinions on different (typically correlated) topics, further demonstrating the emergence of opinion consensus, polarisation and ideological phases. Since its conception, variants of the model of Baumann et al. \cite{Baumann20,Baumann21} have been applied to recommendation algorithms in order to combat opinion polarisation \cite{Santos21}, and explore the addition of repulsion between dissimilar opinions \cite{Cui23}. Focusing on static network topologies, Baumann et al. \cite{Baumann20(2)} extended Taylor's original work \cite{Taylor68} by introducing stubborn agents to a variant of the linear diffusion Abelson model, exploring the role that single, and multiple stubborn agents of differing opinions, can have on societal consensus under a range of network topologies. Acemoglu et al. \cite{Acemoglu13} also explored the role of stubborn agents in a model with variable trust between stochastically interacting agents. Inspired by the Kuramoto model of phased oscillators, Pluchino et al. defined an \textit{opinion changing rate} model for networked agents \cite{Pluchino05}, tested their model on a variety of graph topologies \cite{Pluchino06}, and compared outputs to existing models in the literature \cite{Pluchino06_2}. In a novel application, Giraldo and Passino \cite{Giraldo16} considered a task completion model which simulated a team of individuals who are both attracted and repulsed from each other based on how they perceived other members of the team were accomplishing the given task. The model displayed dynamics that were consistent with behaviours observed in human groups. Leonard et al. \cite{Leonard21} proposed a simple nonlinear model that demonstrated that polarisation in the USA's political system arises due to positive feedback mechanisms of its processes (ideological sorting, faster news cycle, etc.), even going so far as suggesting that the Republicans have crossed a critical irreversible threshold.

Other modelling paradigms besides continuous differential equations have been used to explore opinion dynamics. Friedkin and Johnson \cite{Friedkin90} employed a discrete approach to model opinion evolution when subject to exogenous factors and others' opinions in their interpersonal network. Though the authors found that their work resembled previous models of opinion formation such as that of Wagner \cite{Wagner78}, they noted that their model did not have a close resemblance to Abelson's \cite{Abelson64} original work. Hegselmann and Krause \cite{Hegselmann02} considered the Friedkin-Johnson model alongside similar discrete models lacking the exogenous factors, demonstrating wide varieties of behaviours between global consensus and polarisation, and Milli \cite{Milli21} furthered this by exploring the model in the presence of stochastic noise. Variants of the Friedkin-Johnson model have been applied to understand consensus formation in the 2015 Paris Agreement on climate change \cite{Bernardo21}, and explore/optimise social network topologies that minimise opinion polarisation \cite{Musco18, Zhu21}. Using an agent-based paradigm, Axelrod \cite{Axelrod97} considered agents on a geographical grid who either adopted, or rejected, features of their neighbours over discrete time steps. By grouping culturally aligned agents into nation states, Axelrod showed that the number of stable nations decreases with the increase of the number of features under consideration by agents and the range of interaction, while the number increased with the size of the overall geography up to a critical point and then decreased. It is noteworthy that, when summarising seven previously proposed mechanisms for why consensus isn't a global outcome, Axelrod identified that they overlook the fact people are more likely to interact with those similar to them, which we consider later in our selection of networks. Deffuant et al. \cite {Deffuant00} applied a mixed discrete/continuous time approach that modelled pairs of agents interacting at randomised time intervals, adjusting their respective opinions based on a global threshold. The authors observed critical values of the global threshold that led to the formation of isolated opinion clusters, deviating from global consensus. Lanchier and Mercer \cite{Lanchier24} introduced agent stubbornness to the Deffuant model, which saw a final outcome of global consensus disappear. 

Thus, the topic of opinion dynamics has been quantitatively studied through a number of lenses. Previous works have explored controversialness, and its effect on opinions, in the context of stochastically dynamic networks \cite{Baumann20, Baumann21, Cui23}, leading to emerging echo-chamber behaviours. The novelty of our work is the focus on the effect of controversy, in combination with explicit control over attraction and repulsion topologies. Coupled with the new clustering algorithms offered in this work, this model enables exploration and understanding of the effect that trolls who employ controversy, in a targeted (or untargeted) manner, can have on the opinion profile of a social network.  

The next section defines the model, detailing the application of distinct topologies for attraction and repulsion. Sections \ref{sec:identical} and \ref{sec:different} demonstrate model behaviour for identical, and distinct, attraction and repulsion topologies, respectively. Section \ref{sec:deeper} details model behaviour through the lens of novel clustering algorithms presented in this work, and Section \ref{sec:conclusion} offers conclusions.

\section{Model definition} \label{sec:modeldef}
Consider $N\in\mathbb{Z}$ interacting agents whose opinions are denoted by dynamic variables
\begin{equation}
x_i \in \mathbb{R}, \;\; i \in \{1, \dots, N\}.    
\end{equation}
The opinions of each of the $N$ agents are affected by the interaction of two mechanisms: attraction and repulsion. In its most general form, the networked model which combines both the attraction mechanism of Abelson \cite{Abelson64} with the repulsion mechanism of Baumann \textit{et. al} \cite{Baumann20,Baumann21} is given via
\begin{eqnarray}
    \dot{x}_i = - \frac{1}{N}\left[\sum^N_{j =1} {\cal A}_{ij} (x_i-x_j) -  \sum^N_{j=1}  {\cal R}_{ij} \tanh \alpha (x_i - x_j)\right], \;\; i \in \{1,\dots,N\}, \;\; \alpha \in \mathbb{R}_+.
    \label{general}
\end{eqnarray}
Notably, Equation (\ref{general}) generally consists of two networks ${\cal A}$ and ${\cal R}$, referred to as the \textit{attraction} and \textit{repulsion} \textit{networks}, respectively. The first term in Equation (\ref{general}) is a linear diffusion term which attracts the opinions of agents $i$ and $j$ if they are connected via ${\cal A}$. The second term in Equation (\ref{general}) acts to separate the opinions of agents $i$ and $j$ if they are connected via network ${\cal R}$. The responsiveness of the non-linear hyperbolic tangent term is controlled by the parameter $\alpha$; for small values the responsiveness is small, leading to limited repulsion between agent opinions, whereas large values leads to a strong repulsion between agent opinions. Nevertheless, following \cite{Jayles17}, the functional form of hyperbolic tangent means that the strength of the repulsion of opinions is capped. Following \cite{Baumann20, Baumann21}, we refer to $\alpha$ as \textit{controversialness} as it serves as the control parameter of the repulsion of agent opinions. The symmetry of ${\cal A}$ and ${\cal R}$, combined with the odd functional forms of $x_i-x_j$ and the hyperbolic tangent means that the sum of opinions in Eq.(\ref{general}) is conserved:
\begin{equation}
    \sum^N_{i=1}\dot{x}_i = 0, \;\; \Rightarrow  \sum^N_{i=1}x_i = \sum^N_{i=1}x_i(0) \;\; \forall \;\; t.
    \label{conserved}
\end{equation}

\subsection{Guaranteed fixed points}
It is possible to show that system fixed points exist for all values of $\alpha$. To obtain this result we express Eq.(\ref{general}) in the form of a potential function
\begin{eqnarray}
     \dot{x}_i = - \frac{\partial}{\partial x_i} V(\mathbf{x}), \;\; i \in \{1,\dots,N\},\nonumber\\
     \textrm{where} \;\;V(\mathbf{x})= \frac{1}{N}\sum^N_{\genfrac{}{}{0pt}{}{i,j=1}{i<j}}\left[ \frac{1}{2}{\cal A}_{ij} (x_i-x_j)^2 -\frac{1}{\alpha} {\cal R}_{ij}  \ln \cosh \alpha (x_i - x_j) \right] . \label{potential2}
\end{eqnarray}
and $\mathbf{x} \equiv \{x_1, \dots, x_N\}$. The potential $V(\mathbf{x})$ in Eq.(\ref{potential2}) represents a hyperplane in the variables $\mathbf{x}$, whose local minima give the fixed points of Eq.(\ref{general}). To show that local minima always exist in Eq.(\ref{potential2}) we note that for all $k \in \{1,\dots, N\}$, every variable $x_k$ in Eq.(\ref{potential2}) possesses the limits
\begin{eqnarray}
\lim_{x_k \rightarrow 0} V(\mathbf{x}) &=&\frac{1}{N}\sum^N_{\genfrac{}{}{0pt}{}{j=1}{\ne k}}  \left[ \frac{1}{2}{\cal A}_{kj} x^2_j - \frac{1}{\alpha}{\cal R}_{kj} \ln \cosh \alpha x_j \right] < \infty \nonumber \\
    \lim_{x_k \rightarrow \pm \infty} V(\mathbf{x}) &\sim& \frac{1}{2N} d_k x^2_k,
\end{eqnarray}
where the \textit{degree} $d_k$ is the number of connections to node $k$. Due to $V(\mathbf{x})$ being a continuous, differentiable function in $\mathbb{R}^N$, the mean value theorem guarantees that there must exist at least one local minima in $V(\mathbf{x})$.

\subsection{Order parameter}
In order to measure opinion cohesiveness over all agents, we apply the following \textit{order parameter}
\begin{equation}
    r = \sum^N_{i,j = 1}\frac{\left( \Delta_{i,j}\right)^2}{N^2} \;\;\;\textrm{where}\;\;\; \Delta_{i,j} = | x_i(t_{max})-x_j(t_{max})|
    \label{order}
\end{equation}
which is a normalised measure of the distance between the opinions of all agents at the model's end time (labelled $t_{max})$. Notably, unlike order parameters associated with oscillator models \cite{Strogatz00}, Eq.(\ref{order}) is not bounded on the circle. This will become important in understanding the interplay between network topologies and controversialness on the spread of opinions. 

\subsection{Classification method} \label{sec:classification}


Though the order parameter $r$ detects the phase transition from perfect consensus as a function of $\alpha$, there are 2 limitations which are addressed in this work. Firstly, $r$ is strongly skewed by the range of final opinions (noting that these are not bounded), and secondly, $r$ cannot convey the finer features of the final opinion distribution. The first issue can be mitigated by normalising final opinions, and accounting for outliers, before calculation of $r$. The final opinions are first filtered to remove outliers using a threshold of 1.5*IQR (interquartile range) from the lower ($Q_1$) and upper ($Q_3$) quartiles. If the most extreme opinions are beyond $\pm1$, these opinions are scaled symmetrically on the range $[-1,1]$, such that an opinion at zero is untouched, and the skew of the normalised results matches the skew of the raw results. 

Addressing the second issue, for a more complete understanding of the final opinion distribution, we consider the framework from Devia and Giordano \cite{Devia22} that qualitatively categorises opinion distributions for discrete data. They classified opinion distributions into one of 5 paradigms: perfect consensus, consensus, polarisation, clustering and dissensus. However, we observed that discretising the final opinion distributions introduced artificial divides in the middle of adjacent opinions, therefore distorting the shape. To counter this, we offer a continuous variant of the framework that employs a one-dimensional generalisation of density-based clustering \cite{Kriegel11}, which we use to classify the continuous filtered and normalised final opinion distribution into the same paradigms used by Devia and Giordano. 

Recalling that the separation between final opinions of nodes $i$ and $j$ is defined in Eq.(\ref{order}), we additionally define a \textit{sequence} and a \textit{cluster}:

\begin{worddefs}
  \item[Sequence.] A group of consecutive opinions that are each separated by $\Delta_{i,i+1} \le Th$ (a chosen threshold). 
  \item[Cluster.] A sequence of 5 or more opinions with all $\Delta_{i,i+1} \le$ 0.05 (2.5\%) and the total sequence width $<$ 0.5 (25\%).
\end{worddefs}
We define the following criteria for each paradigm, with thresholds approximately calibrated to match the parameters they used for discrete opinions, though we define 2 different ways to identify both clustering and dissensus. An opinion distribution may satisfy the criteria for multiple paradigms (e.g. consensus and polarisation), but our analysis assigns a primary paradigm according to the order in the list below.

\begin{enumerate}
    \item \textit{Perfect consensus}: $\ge$ 50\% of opinions are in one sequence with all $\Delta_{i,i+1}\le$ 0.01 (0.5\%).
    \item \textit{Consensus}: $\ge$ 50\% of opinions are in one cluster. 
    \item \textit{Polarisation}: at least one $\Delta_{i,i+1} > 0.5$ (25\%).
    \item \textit{Clustering (type A)}: at least 2 clusters and $\ge$ 50\% of opinions are within clusters.
    \item \textit{Clustering (type B)}: 2 or more $\Delta_{i,i+1} > 0.2$ (10\%) --- derived from the definition presented by Devia and Giordano \cite{Devia22}
    \item \textit{Dissensus (dispersion)}: $\ge$ 50\% of opinions are in one sequence with all $\Delta_{i,i+1} \le 0.05$ (2.5\%), but the total sequence width $\ge 0.5$ (25\%) --- introduced in addition to the default dissensus used by Devia and Giordano \cite{Devia22} to qualify the specific behaviour observed in continuous opinions.
    \item \textit{Dissensus (by default)}: If none of the above paradigms apply, then by default the paradigm is assigned as dissensus.
\end{enumerate}

\subsection{Network structure}

\begin{figure}[!ht]
    \centering
\includegraphics[width=1\textwidth]{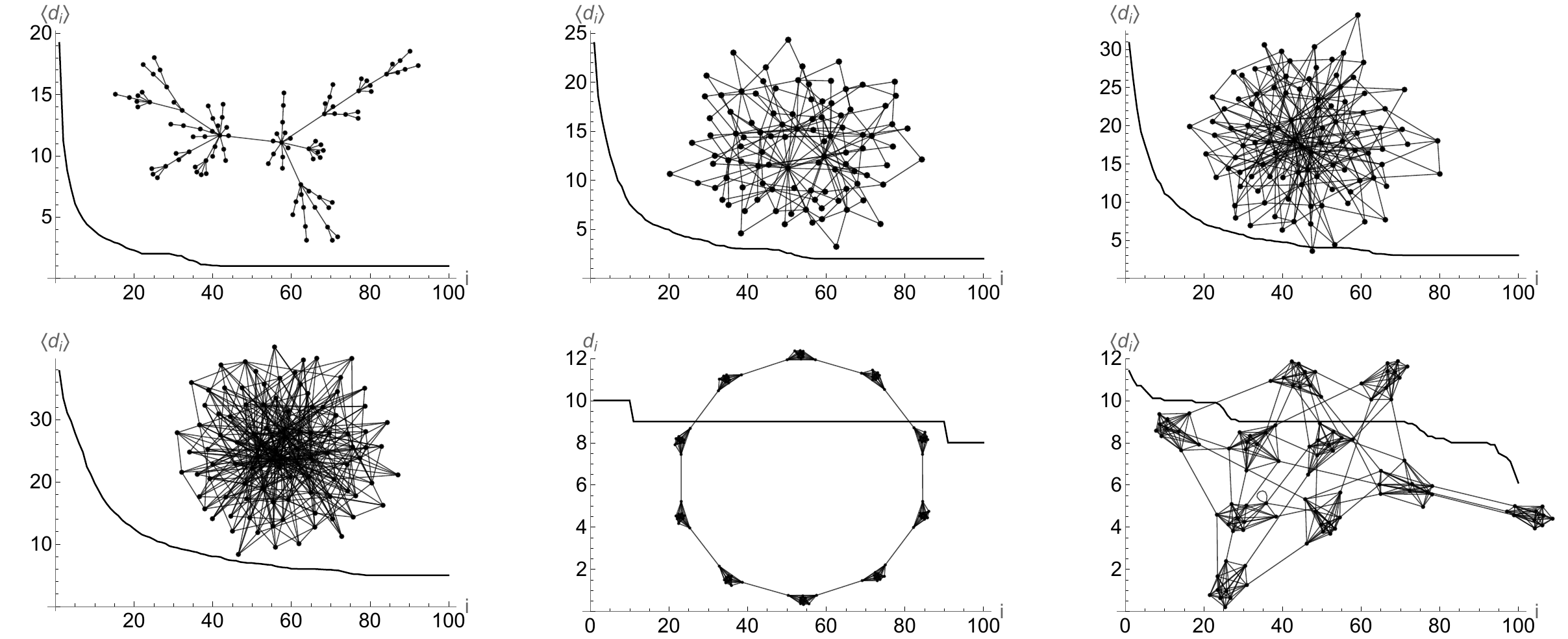}
    \caption{For six types of network, $\{BA_1, BA_2$, $BA_3$, $BA_5, C_C$, $C_R$\}, we show the average degree values $\langle d_i \rangle$, of each node $i$, sorted from highest to lowest, sampled from 25 graph instances. Each panel additionally displays a specific instance of the graph under consideration.}
\label{fig:graphs_ex}
\end{figure}

As the defining Equation (\ref{general}) has only one free parameter $\alpha$, the different forms of the network topologies of ${\cal A}$ and ${\cal R}$ are important parameters to understand and quantify model behaviour. In this work, we consider Barab\'{a}si-Albert ($BA$) networks, connected caveman ($C_C$) networks and relaxed caveman ($C_R$) networks. All networks consist of 100 nodes. Since network topology has a strong effect on the model behaviour, and all networks are generated randomly, statistical results are obtained using either 10 or 25 variants of each graph with recorded seeds (see section \ref{sec:setup}). Figure \ref{fig:graphs_ex} displays examples of 4 $BA_k$ graphs, $C_C$ and $C_R$, along with the average degree of each node $i$ (sorted by decreasing degree) for 25 random variants of the $BA$ graphs, and 10 for $C_C$, generated with fixed seeds.

The $BA$ network model approximates scale-free graphs of preferential attachment, typical of social networks. These networks are generated by adding new nodes each with $k \in \mathbb{Z}_+$ edges attached preferentially to existing high degree nodes --- labelled $BA_k$. $BA_k$ graphs are considered in this work for both ${\cal A}$ and ${\cal R}$. 

Caveman graphs consist of $l$ fully-connected cliques of size $k$. They are an intuitive candidate for ${\cal A}$, representing groups of like-minded people who interact frequently in an opinion-reinforcing feedback loop. As we consider fully connected graphs in this work, the single $C_C$ graph rewires a single edge per clique to an adjacent clique, ensuring that the graph is connected while retaining the densely-connected cliques. $C_R$ graphs randomly rewire each edge with probability $p$ to link different cliques, thus functioning similarly (but with denser cliques) to the stochastic block model used by Baumann et al. \cite{Baumann20(2)}. In all $C_C$ and $C_R$ graphs, the 100 nodes are assigned to 10 cliques of 10 nodes, while all $C_R$ graphs use $p = 0.1$ for rewiring.

\FloatBarrier
\subsection{Experimental set up} \label{sec:setup}

\begin{figure}[!ht]
    \centering
\includegraphics[width=1\textwidth]{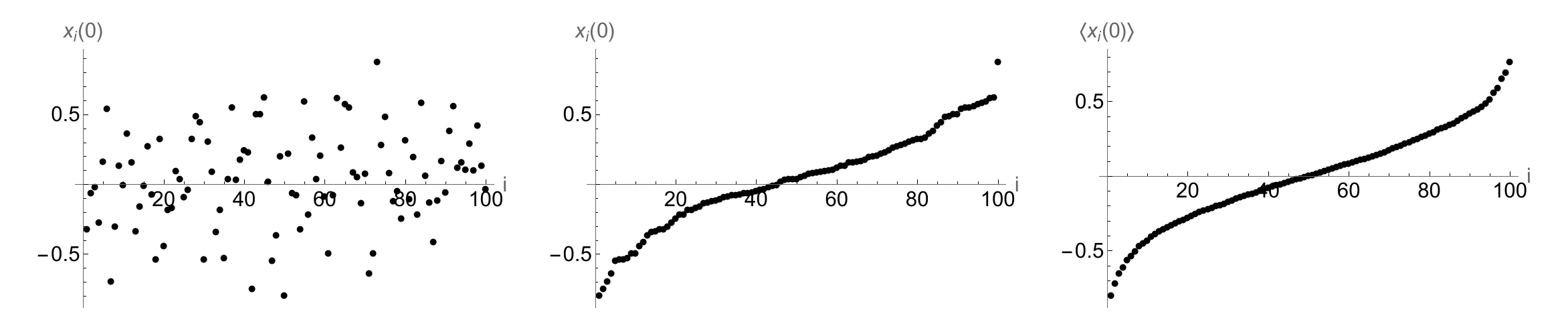}
    \caption{Left panel displays one instance of initial conditions $x_i(t=0)$, $i \in \{1,100\}$, applied in this work, randomly generated from a normal distribution, with standard deviation equal to $\frac{1}{3}$. Middle panel displays the same initial conditions as on the left, now ordered. Right panel displays the ordered average of 25 of the initial conditions considered in this work.}
\label{fig:ini_cond}
\end{figure}

The opinion evolution that occurs through our model in Eq.(\ref{general}) is also sensitive to initial conditions, though typically less so than network structure. To account for this, we iterate over randomised instances of initial conditions, generated from a normal distribution with standard deviation equal to $\frac{1}{3}$, again recording seed numbers to enable reproducibility. In Figure \ref{fig:ini_cond}, the left panel displays one instance of initial conditions, $x_i(t=0)$ for $i \in \{1,100\}$, while the middle panel presents the same initial condition ordered. Finally, the right panel displays the ordered average of 25 instances of initial conditions considered in this work --- converging to a normal distribution.

Three experiments are used to examine the model behaviour in this work, with results of each discussed in its own section:
\begin{itemize}
    \item Section \ref{sec:identical}: The average order parameter is determined for ${\cal A} = {\cal R}$ using networks $\{BA_1,BA_2,BA_3,BA_5,BA_7,BA_{21}, BA_{99}\}$. For each network type, the model is run for 25 instances of the network with 25 initial conditions (625 total combinations), sweeping across 36 values of $\alpha \in [0, 7]$.
    \item Section \ref{sec:different}: The average order parameter is determined for ${\cal A} \ne {\cal R}$ using pairs of networks $\{BA_1,BA_2,BA_3\}$, yielding 9 pairs of network types. For each network combination, the model is run for 10 instances of both ${\cal A}$ and ${\cal R}$ (with different seeds) and with 10 initial conditions (1000 total combinations), sweeping across 26 values of $\alpha$ --- the range of $\alpha$ is chosen independently for each combination.
    \item Section \ref{sec:deeper}: Deeper analysis is undertaken for ${\cal A} \ne {\cal R}$ using 4 different networks for ${\cal A} \in \{BA_1$, $BA_3$, $C_C$, $C_R\}$ and ${\cal R} \in BA_1$. These 4 pairs of networks were selected as they demonstrate the full breadth of behaviour observed in the model. Furthermore, they reflect the intuitive notion that people prefer to interact with like-minded people, and so the ${\cal A}$ graphs (some with topological communities) are denser than the sparse ${\cal R}$ graphs. For these pairs, the same set-up was run as section \ref{sec:different} (1000 total combinations per pair across 26 values of $\alpha$ spanning an independent range). Analysis in this section includes: categorisation of resulting opinion distributions, calculation of normalised metrics, maximum distance between opinions, number of clusters, number of topological communities (defined as connected components after unclustered nodes and inter-cluster links are removed), number of clustered/unclustered/outlying opinions, and topological impact on paradigm.
\end{itemize}

\section{Identical attraction and repulsion networks} \label{sec:identical}
\subsection{Example model behaviour}

\begin{figure}[!ht]
    \centering
\includegraphics[width=1\textwidth]{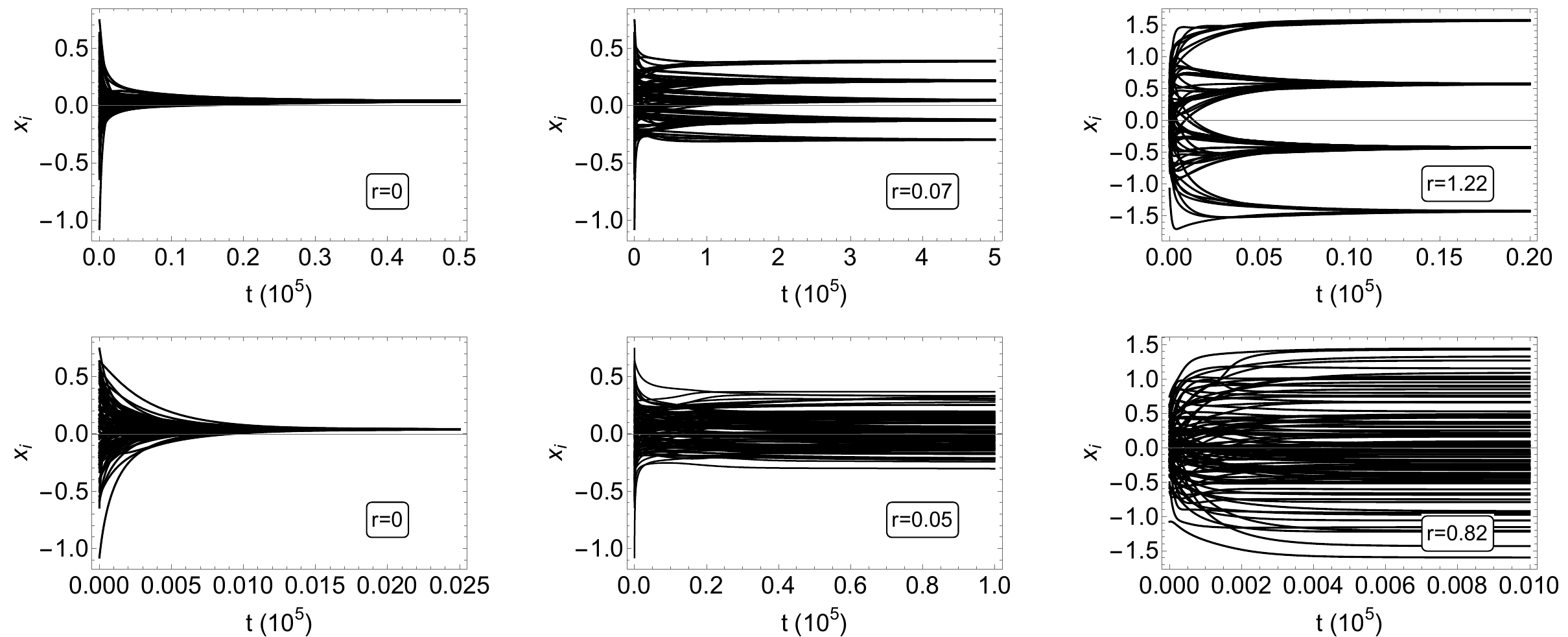}
    \caption{Model output examples for same attraction and repulsion networks (${\cal A} = {\cal R}$) with top and bottom rows showing $BA_1$ and $BA_2$ networks, respectively. Columns 1, 2, and 3 show opinion trajectories for $\alpha$ values of $0.75$, $1.01$, and $5$, respectively. Insets on each panel list the order parameter value of Eq.(\ref{order}) for each model output. All outputs have the same initial conditions.}
\label{fig:example}
\end{figure}

Figure \ref{fig:example} provides example model behaviours for different iterations of ${\cal A} = {\cal R}$. The top and bottom rows show results for $BA_1$ and $BA_2$ networks, respectively. Columns 1, 2, and 3 show opinion trajectories for $\alpha$ values of $0.75$, $1.01$, and $5$, respectively. The panels on the left, for $\alpha=0.75$, show all 100 trajectories converging to the same opinion given enough time. This behaviour notably changes in the middle panels, for $\alpha=1.01$, with all trajectories converging on one of 5 opinion clusters in the top middle panel for $BA_1$, or landing on a spread in the bottom middle panel for $BA_2$. We shall explain this phase transition behaviour --- from all opinions converging for $\alpha \le 1$ to diverging for $\alpha > 1$ --- for ${\cal A} = {\cal R}$ shortly. For larger controversialness ($\alpha =5$) in the right column, the behaviour displayed in both panels is similar to what was exhibited in the middle column, though the larger value of $\alpha$ noticeably pushes apart the final position of opinions, leading to significantly larger values of the order parameter $r$.

\subsection{Critical controversialness value} \label{sec:critical}
Assuming ${\cal A} = {\cal R}$, Equation (\ref{general}) becomes 
\begin{eqnarray}
        \dot{x}_i = - \frac{1}{N}\sum^N_{j =1} {\cal A}_{ij} \left[ (x_i-x_j) -  \tanh \alpha (x_i - x_j) \right], \;\; i \in \{1,\dots,N\}, \;\; \alpha \in \mathbb{R}_+.\label{same}
\end{eqnarray}
This simplification is more amenable to analytical understanding of model behaviour. Exploring the $\alpha < 1$ behaviour of the model, assuming
\begin{eqnarray}
    x_i \approx x_j \;\;\; \Rightarrow \;\;\; \tanh \alpha (x_i - x_j ) \approx \alpha (x_i - x_j ), \;\; \forall \;\; \{i, j \} \label{assump}
\end{eqnarray}
Equation (\ref{same}) becomes
\begin{eqnarray}
        \dot{x}_i \approx - \frac{(1-\alpha)}{N}\sum^N_{j =1} {\cal A}_{ij}  (x_i-x_j), \;\; i \in \{1,\dots,N\}, \;\; \alpha \in \mathbb{R}_+,\label{same2}
\end{eqnarray}
which has the corresponding graph-\textit{Laplacian} form
\begin{equation}
       \dot{x}_i = - \frac{(1-\alpha)}{N}\sum^N_{j=1}{\cal L}^{{\cal A}}_{ij} x_j, \;\; i \in \{1,\dots,N\}, \;\; \alpha \in \mathbb{R}_+.\label{same3}
\end{equation}
Assuming the network ${\cal A}$ is entirely connected, the real-valued eigenvalues of ${\cal L}^{{\cal A}}$ --- labelled $\lambda^{\cal A}_r$ --- are ordered via
\begin{equation}
   0 = \lambda^{\cal A}_0 < \lambda^{\cal A}_1 \le \lambda^{\cal A}_2 \le \dots \le \lambda^{\cal A}_{N-1}.
    \label{spec1}
\end{equation}
Applying the properties of the Laplacian eigenvalues and eigenvectors (refer to \ref{app:lap} for details), the steady-state solution to Eq.(\ref{spec1}) is given as
\begin{equation}
 x_i(t\rightarrow \infty) = \sum^N_{j=1} \frac{x_j(0)}{N} \equiv \bar{x}(0), \;\; \forall \;\; i \in \{1, \dots, N\} \label{steadylin}
\end{equation}
which is the intuitive result that if all opinions are equally weighted with little controversy, then opinions will converge to the average given enough time. 

\begin{figure}[!ht]
    \centering
\includegraphics[width=0.5\textwidth]{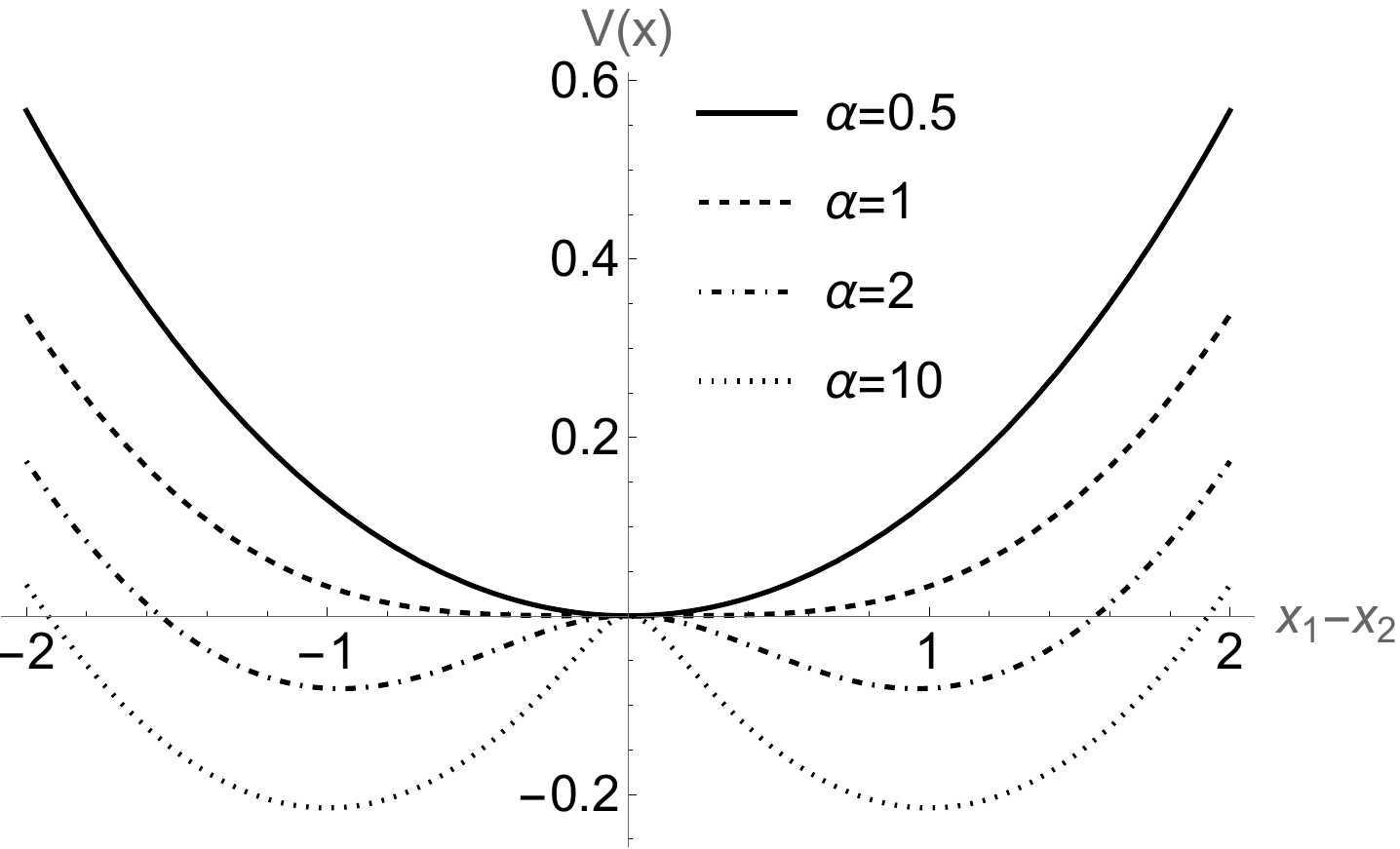}
    \caption{Examples of the potential function in Eq.(\ref{N_2}) for $N=2$ with different values of $\alpha$.}
\label{fig:potential1}
\end{figure} 

Exploring system behaviour beyond $\alpha <1$, Figure \ref{fig:potential1} gives the simplest example ($N=2$) of Eq.(\ref{potential2}) for the system potential:
\begin{equation}
    V(x_1,x_2) = \frac{1}{2}\left[ \frac{1}{2}(x_1-x_2)^2 - \frac{1}{\alpha} \ln \cosh \alpha (x_1 - x_2) \right] \label{N_2}
\end{equation}
Figure \ref{fig:potential1} shows that for $\alpha \le 1$, the fixed point of the system is $x_1 = x_2$, consistent with the result in Eq.(\ref{steadylin}). For $\alpha >1$, two fixed points emerge, with the valid point dependent on the initial conditions. Thus there are only two possible macroscopic behaviours for a given network with ${\cal A} = {\cal R}$: 
\begin{itemize}
    \item{If $\alpha \le 1$, all opinions will converge to the average of the initial conditions.} 
    \item{If $\alpha > 1$, all opinions will diverge by a fixed amount from each of their connected neighbours, 
    potentially forming clusters of non-adjacent nodes with common opinions (as seen in the middle and right columns of Figure \ref{fig:example}).}
\end{itemize}
In dense networks this distribution becomes less predictable, nonetheless adjacent nodes are unable to maintain identical opinions. Thus for ${\cal A} = {\cal R}$ and $\alpha >1$, clusters of opinions may appear to emerge macroscopically, nevertheless they are an artifact of alternating discrete opinions rather than a tight community collectively forming an opinion. This is behaviour is explicitly demonstrated in \ref{app:star}. 

\subsection{Statistical analysis of general graphs}
\begin{figure}[!ht]
    \centering
\includegraphics[width=0.5\textwidth]{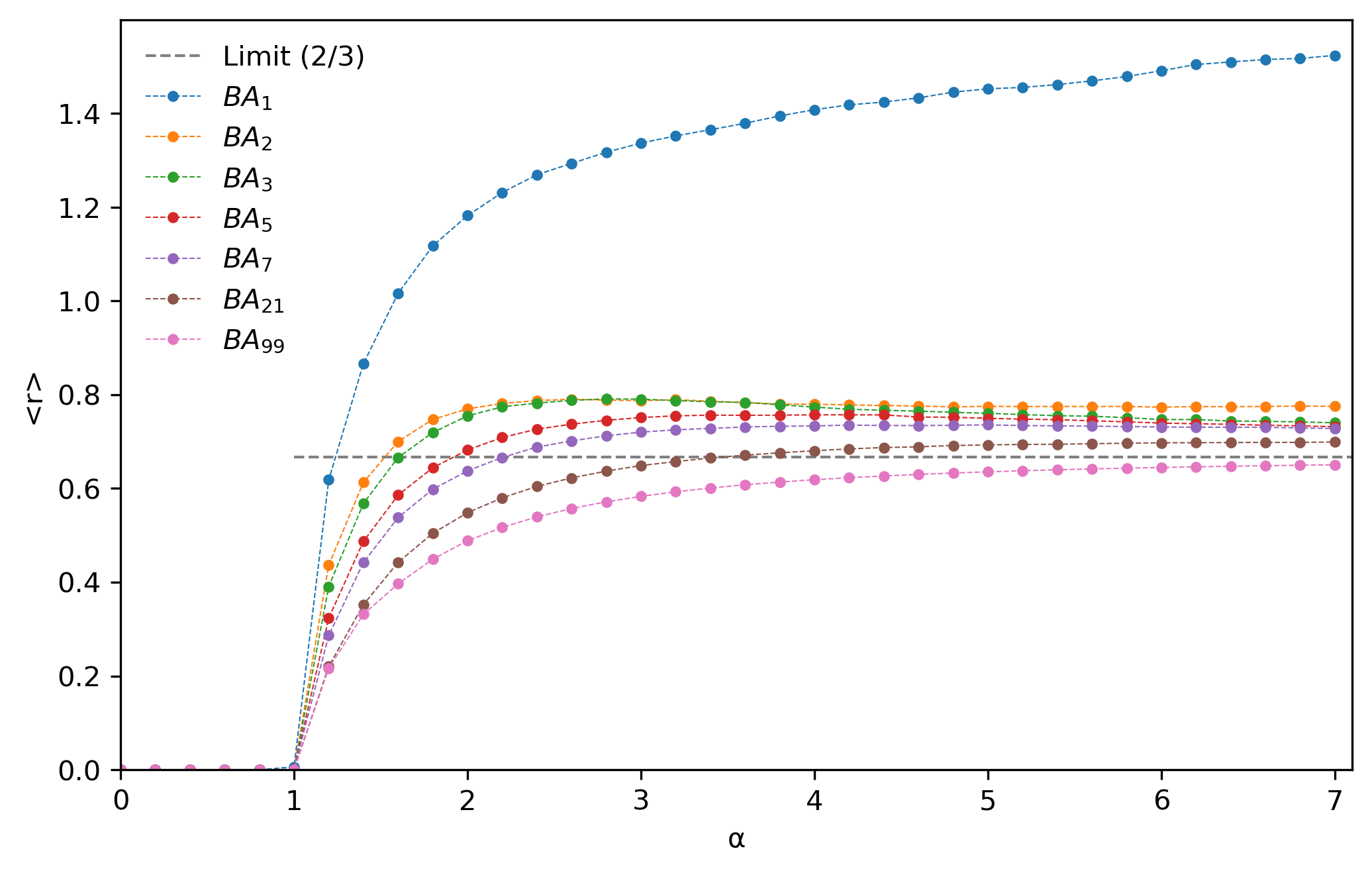}
    \caption{Average results of the order parameter for increasing $\alpha$ for different classes of $BA$ networks, plotted with the theoretical limit of the all-to-all network ($r=2/3$) given in Eq.(\ref{theor}).}
\label{fig:ARequal2}
\end{figure}

Figure \ref{fig:ARequal2} presents order parameter results for each of the different classes of $BA$ networks considered in this work. For an analysis of the most extreme case of $BA_1$ --- the \textit{star-graph} --- refer to \ref{app:star}. The order parameter values $\langle r \rangle$ are obtained by collecting values for Eq.(\ref{order}) over 25 instances of $BA_k$ network, in addition to 25 initial condition instances (a total of 625 data points per $\alpha$-value). The trajectory gives the average value of $r$ per $\alpha$-value. Notably, since the attraction and repulsion graphs are identical, the order parameter is zero for $\alpha <1$, as per the analysis in section \ref{sec:critical}. 
We do not present error bars for the statistical results of this case due to ${\cal A}= {\cal R}$ restricting output variability compared to ${\cal A} \ne {\cal R}$.

Additionally, it is possible to obtain an exact expression for the order parameter value for the all-to-all network with $\alpha \gg 1$, given via
\begin{eqnarray}
    r = \frac{2}{3}\left(1-\frac{1}{N^2} \right) \approx \frac{2}{3},
    \label{theor}
\end{eqnarray}
with the details of the derivation of Eq.(\ref{theor}) given in \ref{sec:theo}. Figure \ref{fig:ARequal2} demonstrates the validity of Eq.(\ref{theor}) for large values of $\alpha$ in the all-to-all network, with the $BA_{99}$ order parameter settling on the value $r=2/3$ as $\alpha$ becomes large. 

\section{Different attraction and repulsion networks} \label{sec:different}
\subsection{Example model behaviour}

\begin{figure}[!ht]
    \centering
\includegraphics[width=1\textwidth]{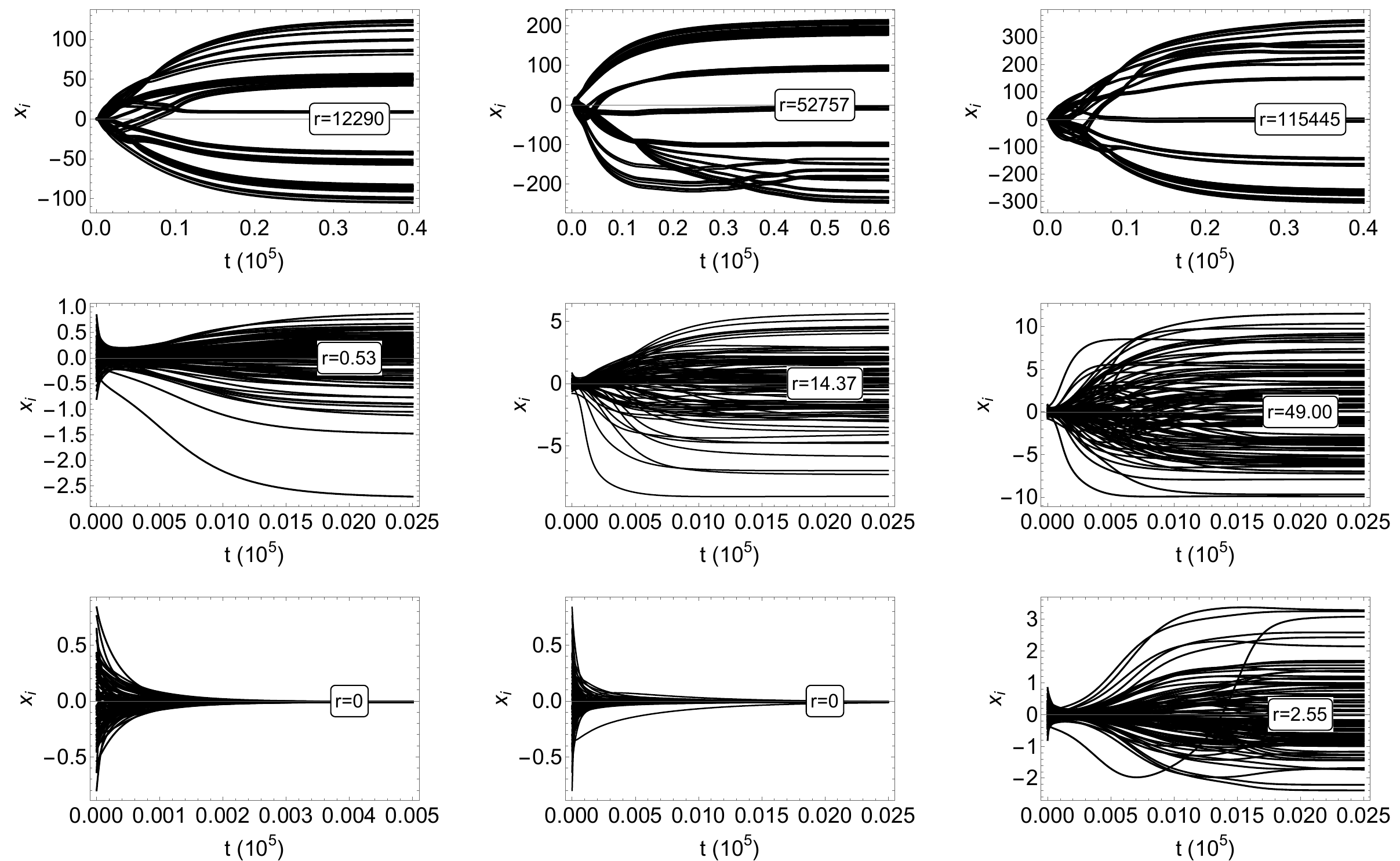}
    \caption{Model output examples for different attraction and repulsion networks. Top, middle, and bottom rows show outputs for attraction networks derived from $BA_1$, $BA_2$, and $BA_3$, respectively. Left, middle, and right columns show outputs for repulsion networks derived from $BA_1$, $BA_2$, and $BA_3$, respectively. Insets on each panel list the order parameter value of Eq.(\ref{order}) for each model output. All outputs have $\alpha=0.25$, and employ the same initial conditions.}
\label{fig:example2}

\end{figure}
Figure \ref{fig:example2} presents model outputs for the case of ${\cal A} \ne {\cal R}$, with consistent initial conditions, and $\alpha =0.25$ for each panel. The top, middle, and bottom rows show outputs for attraction networks derived from $BA_1$, $BA_2$, and $BA_3$, respectively. Correspondingly, the left, middle, and right columns show outputs for repulsion networks derived from $BA_1$, $BA_2$, and $BA_3$, respectively. The top row, for ${\cal A} \in BA_1$, shows trajectories similar to the top middle and right of Figure \ref{fig:example}, where opinions form clusters, indicating that relatively low connectivity of the attraction networks leads to clustering of opinions. Nevertheless, the top row in Figure \ref{fig:example2} has significantly larger relative distances between each of the opinion clusters, with an $r$-value 5--6 orders of magnitude greater than anything seen in Figure \ref{fig:example}, indicating that the case of different attraction and repulsion networks leads to a larger range of final opinion values, even for lower controversialness. The middle and bottom rows display a significantly smaller range for the final opinions due to more connections in the attraction network. Focusing left-to-right, the addition of more connections in the repulsion networks has an equivalent effect to increasing the controversialness parameter, which further distances the final opinion values of all agents. 

\subsection{Statistical analysis of general graphs}
\begin{figure}[!ht]
    \centering
\includegraphics[width=1\textwidth]{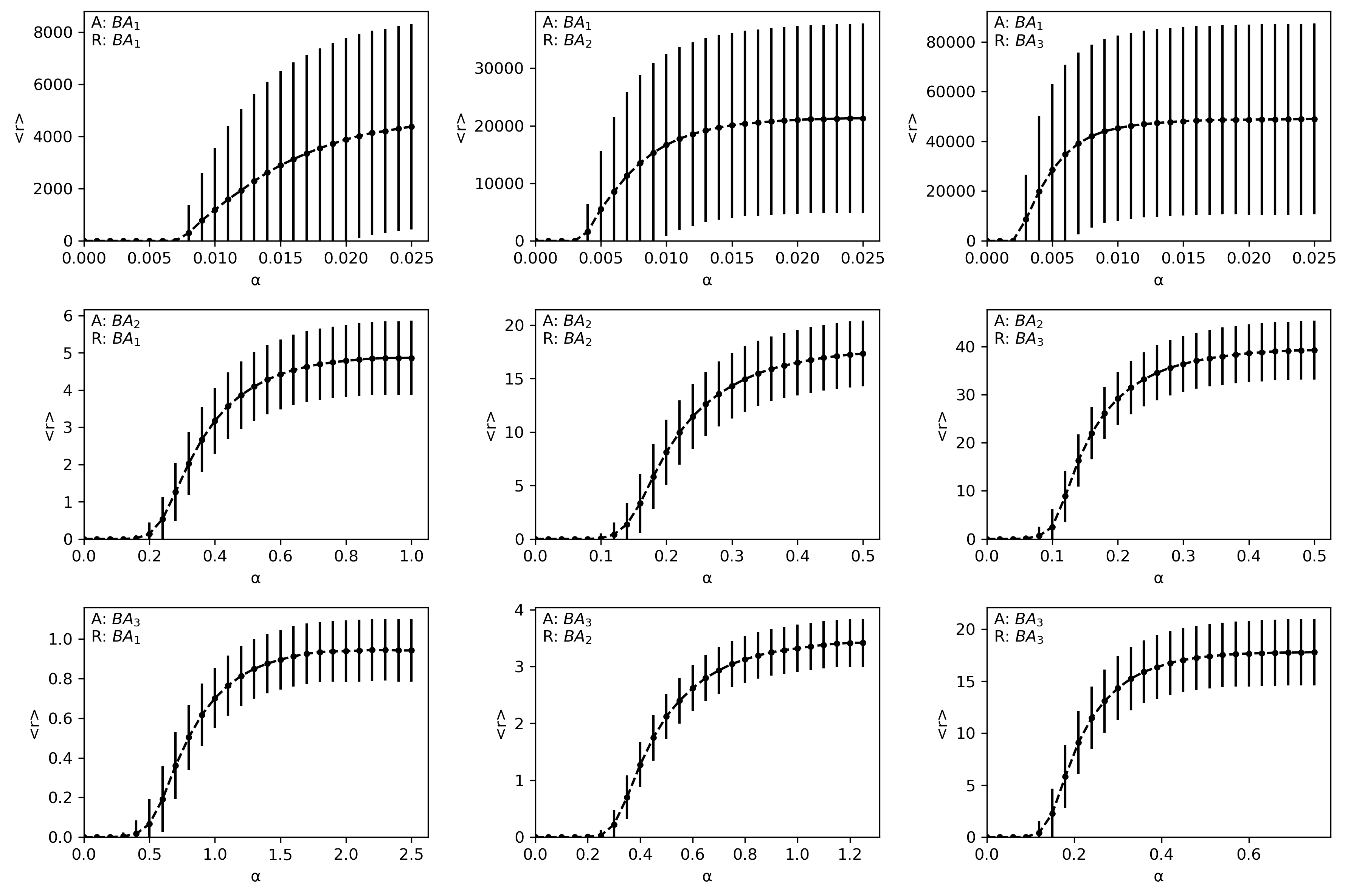}
    \caption{Average results (with error bars representing the standard deviation) of the order parameter for increasing $\alpha$ for different classes of BA networks. Top, middle, and bottom rows show outputs for attraction networks derived from $BA_1$, $BA_2$, and $BA_3$, respectively. Left, middle, and right columns show outputs for repulsion networks derived from $BA_1$, $BA_2$, and $BA_3$, respectively.}
\label{fig:ARdiff}
\end{figure} 

Figure \ref{fig:ARdiff} presents order parameter results for each of the different combinations of $BA$ networks when attraction and repulsion graphs are different. The order parameter values $\langle r \rangle$ are obtained by collecting values for Eq.(\ref{order}) over 10 instances of $BA_k$ graph for both attraction and repulsion networks, in addition to 10 initial condition instances (a total of 1000 data points per $\alpha$-value). The trajectory gives the average value of $r$ per $\alpha$-value, with the error bars denoting the standard deviation over the attraction and repulsion $BA_k$-graphs, and initial condition combinations.

Comparing Figures \ref{fig:example} and \ref{fig:ARequal2} for ${\cal A} = {\cal R}$, with Figures \ref{fig:example2}--\ref{fig:ARdiff} for ${\cal A} \ne {\cal R}$, we see that finding controversy in the opinions of others who we have little relation to does considerably more to drive apart opinions in a social network than if we find controversy in the opinions of others who we also have innate connection to. This phenomenon was displayed in the 2012 Assam riots, caused by social media trolls inflaming Hindu-Muslim religious-ethnic tensions in northern India, leading to hundreds of thousands of refugees fleeing the area \cite{Goolsby13}. On a geopolitical level, state sponsored trolls, such as those from the Russian Internet Research Agency \cite{Dawson19}, encourage controversy by spreading mis/disinformation between those sitting on different ideological viewpoints \cite{Kurowska18}. The outputs of the model, especially when comparing ${\cal A} = {\cal R}$ and ${\cal A} \ne {\cal R}$, enables quantitative appreciation of the effect that trolls can potentially have when sowing division between members of society who have little reason or opportunity to interact outside of heated online ideological arguments \cite{Mochon24}. Though the current analysis gives a sense of scale of the changes brought about due to differing toplogies, in the next Section we shall offer cluster detection algorithms which demonstrate these changes in greater detail.

\subsection{Critical controversialness value}
\begin{figure}[!ht]
    \centering
\includegraphics[width=1\textwidth]{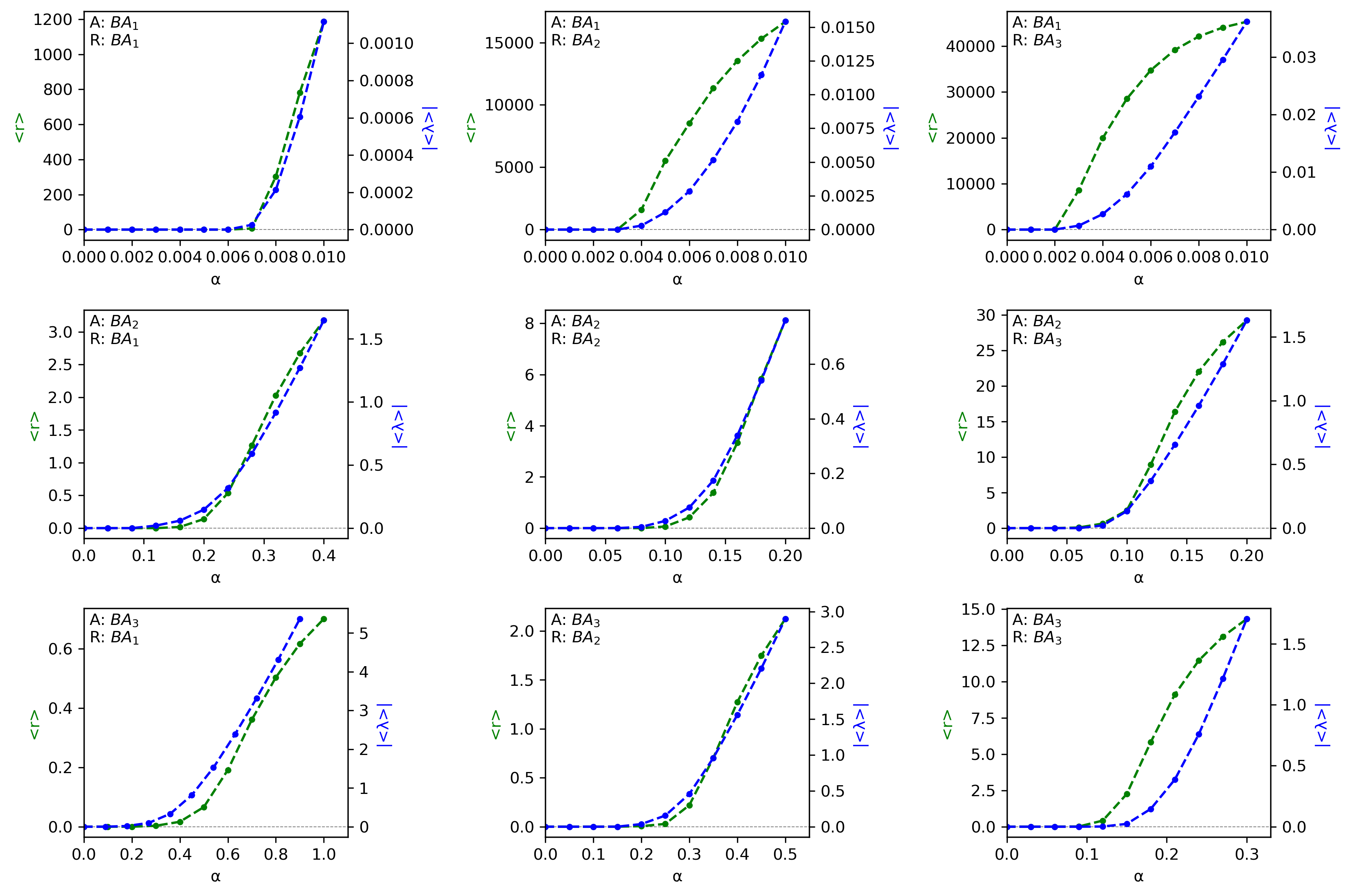}
    \caption{Average value of the modulus of the smallest eigenvalue of ${\cal H}(\alpha)$ (blue), as a function of $\alpha$, for all the 100 combinations of the attraction and repulsion graphs used to generate Figure \ref{fig:ARdiff}. Additionally, the average order parameter given in Figure \ref{fig:ARdiff} has been reproduced (green) for convenience.}
\label{fig:metric_eigen}
\end{figure}

The defining system for ${\cal A} \ne {\cal R}$ in Eq.(\ref{general}) is of course more complicated than Eq.(\ref{same}) for ${\cal A} = {\cal R}$. In order to explore the critical $\alpha$ values for Eq.(\ref{same}) which see opinions deviate from complete consensus, we again assume Eq.(\ref{assump}) for all $\mathbf{x}$ values, which sees Eq.(\ref{general}) become
\begin{equation}
       \dot{x}_i = - \frac{1}{N}\sum^N_{j=1}\underbrace{\left({\cal L}^{{\cal A}}_{ij} - \alpha {\cal L}^{{\cal R}}_{ij}\right)}_{\equiv \;{\cal H}_{ij}(\alpha)}  x_j, \;\; i \in \{1,\dots,N\}, \;\; \alpha \in \mathbb{R}_+.\label{diff3}
\end{equation}
The real-valued symmetric matrix ${\cal H}(\alpha)$ determines the interaction of the the opinions ${\mathbf{x}}$ in the linear regime. Unlike the Laplacian ${\cal L}^{\cal A}$ in Eq.(\ref{same3}) whose eigenvalues in Eq.(\ref{spec1}) are assured to be positive semi-definite, the eigenvalues of matrix ${\cal H}(\alpha)$, labelled $\lambda^{\cal H}$, have weaker properties for general values of $\alpha$, namely
\begin{equation}
    0 \ge \lambda^{\cal H}_0 < \lambda^{\cal H}_1 \le \lambda^{\cal H}_2 \le \dots \le \lambda^{\cal H}_{N-1}.
\end{equation}
In the linear regime, the solution to Eq.(\ref{diff3}) exists as exponentiated eigenvalues --- refer to \ref{app:lap} for more details. Hence, we expect the system in Eq.(\ref{diff3}) to exhibit stability and dynamically decay to the average of the initial opinions if the smallest eigenvalue of ${\cal H}(\alpha)$, labelled $\lambda^{\cal H}_0$, equals zero. To test this assertion, Figure \ref{fig:metric_eigen} shows the average value of the modulus of $\lambda^{\cal H}_0$, as a function of $\alpha$, for all the 100 combinations of ${\cal A}$ and ${\cal R}$ used to generate the ensemble order parameter averages given in Figure \ref{fig:ARdiff}.

Each panel in Figure \ref{fig:metric_eigen} shows that for small enough $\alpha$, $\lambda^{\cal H}_0$ is equal to zero. Moreover, when plotted alongside the reproduced order parameter curves from Figure \ref{fig:metric_eigen}, it is apparent that $\lambda^{\cal H}_0$ begins to deviate from zero (becoming negative) at approximately the same $\alpha$ values that we see the order parameters deviate from zero --- indicating that the linear assumption in Eq.(\ref{assump}) is no longer valid. This demonstrates that the linear system in Eq.(\ref{diff3}) is able to capture the critical $\alpha$ values which lead to the breaking of global consensus of opinions.

\section{Categorisation analysis} \label{sec:deeper}

\subsection{Example model behaviour}

\begin{figure}[!ht]
    \centering
\includegraphics[width=1\textwidth]{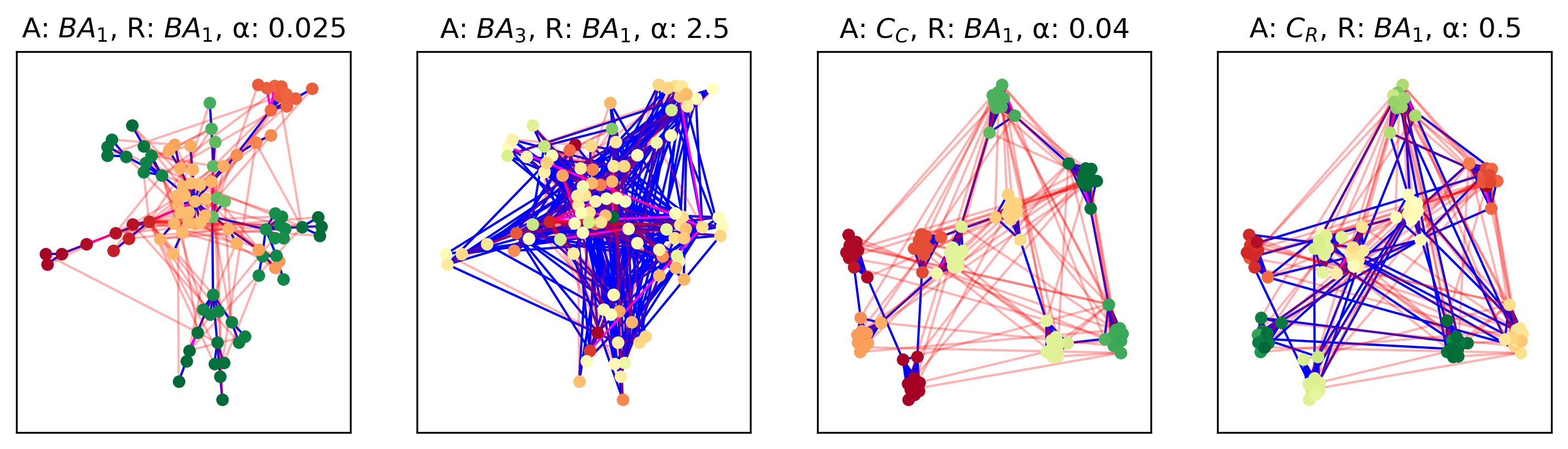}
    \caption{Example networks for ${\cal A}$ graphs $BA_1$, $BA_3$, $C_C$ and $C_R$, paired with ${\cal R} \in BA_1$. The nodes are coloured using a continuous colour map based on the final opinions for each node after the model was run for $10^5$ time units and the maximum $\alpha$ values used for the corresponding sweep. ${\cal A}$ links are blue, ${\cal R}$ links are red.}
\label{fig:4_example_nets}
\end{figure}

Examples of the 4 ${\cal A}$ graphs paired with ${\cal R} \in BA_1$ are shown in Figure \ref{fig:4_example_nets}. The nodes are coloured using a continuous colour map based on the final opinions for each node after the model was run for $10^5$ time units. 
A consistent spring layout has been chosen for the $BA_1$ and $BA_3$ graphs (left), in addition to the $C_C$ and $C_R$ graphs (right), highlighting the similarities and differences within each pair. Notably, for $BA_1$, $C_C$ and $C_R$, groups of nodes that are topologically linked end up with similar opinions, whereas the $BA_3$ example shows a seemingly random distribution of node colours.

\begin{figure}[!ht]
    \centering
\includegraphics[width=1\textwidth]{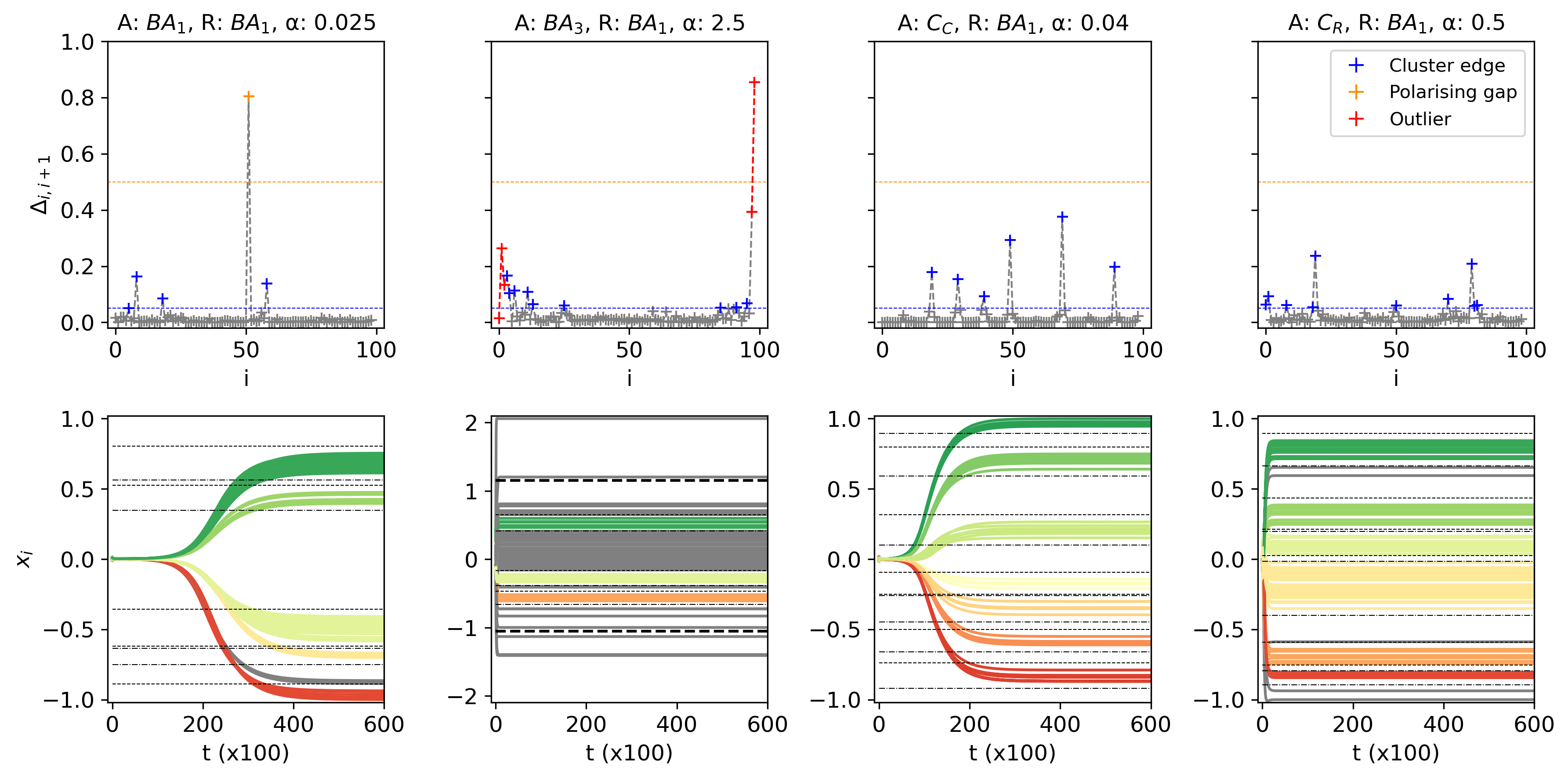}
    \caption{$\Delta_{i,i+1}$ plots (top row) and normalised results (bottom row) for the 4 example networks in Figure \ref{fig:4_example_nets}, with the model run for $10^5$ time units. 
    For $\Delta_{i,i+1}$ plots, nodes are sorted by increasing final opinion and shown as coloured crosses: red (outliers), orange ($>Th_{pol}$), blue ($>Th_{cl}$) and grey ($<Th_{cl}$). For opinion trajectories, individual curves are coloured based on the cluster the final opinion is assigned to, with grey lines representing outliers, or opinions not assigned to a cluster. Overlayed on the trajectories are thick dashed lines representing the cutoff for outliers, and thin dashed/dashdot lines representing the upper/lower limits for each cluster. Each example is classified as the most prevalent paradigm from the corresponding experiment, respectively: polarisation, dissensus, clustering and clustering.}
\label{fig:4_example_results}
\end{figure}

We illustrate our density-inspired clustering approach based on each $\Delta_{i,i+1}$ pair in Figure \ref{fig:4_example_results}, using the same example networks and model results (after normalisation). The top panels plot $\Delta_{i,i+1}$ against node indices, sorted by lowest to highest final opinions. The thresholds for polarisation ($Th_{pol} = 0.5$), and where a cluster ends ($Th_{cl} = 0.05$), are represented with dashed horizontal lines (orange and blue respectively). Crosses representing each $\Delta_{i,i+1}$ are coloured red if the node is an outlier (as defined in section \ref{sec:classification}), orange if greater than $Th_{pol}$, blue if greater than $Th_{cl}$, and grey if below $Th_{cl}$. Since no example is provided for perfect consensus, $\Delta_{i,i+1}$ below this threshold ($Th_{pc} = 0.01$) are not coloured here. 
The bottom row shows opinion trajectories over $6\times 10^4$ time units. Where opinions are clustered, the lines are coloured discretely based on the corresponding cluster, and the upper (max. opinion + 0.05) and lower (min. opinion - 0.05) bounds of the cluster are shown by horizontal dashed and dash-dot lines; unclustered or outlier opinions are coloured grey, and the thresholds for outliers (1.5*IQR from $Q_1$/$Q_3$) are shown with thick horizontal dashed lines (only seen for the ${\cal A} \in BA_3$ example).

For ${\cal A} \in BA_1$, there is a single $\Delta_{i,i+1} > Th_{pol}$, corresponding to the large gap between the upper band and lower band of opinions in the trajectories; leading to the polarisation paradigm, though the algorithm also identifies clusters within the 2 polarised bands. Conversely, for ${\cal A} \in BA_3$, though there is a sequence of more than 50 $\Delta_{i,i+1} < Th_{cl}$, the total width of this sequence is $>$0.5 and thus not counted as a cluster, resulting in an assigned paradigm of dissensus (dispersion). There are also several outliers separated by high $\Delta_{i,i+1}$ values, as well as multiple instances of $Th_{pol}>\Delta_{i,i+1}>Th_{cl}$; but as some of these sequences contain less than 5 nodes, the number of clusters is less than the number of $Th_{pol}>\Delta_{i,i+1}>Th_{cl}$ instances. 3 clusters of 5+ nodes are detected, though they are not visually distinct without the colouring. 
For ${\cal A} \in C_C$, there are pronounced $\Delta_{i,i+1}$ sequences of very small width, broken up by individual $Th_{pol}>\Delta_{i,i+1}>Th_{cl}$ instances, corresponding to 7 opinion clusters --- classified as the clustering paradigm. Finally, for ${\cal A} \in C_R$, though the instances of $Th_{pol}>\Delta_{i,i+1}>Th_{cl}$ are not as regular, and the clusters aren't as tight (with several opinions unclustered), there are 6 clusters with the assigned paradigm still clustering. 

\begin{figure}[!ht]
    \centering
\includegraphics[width=1\textwidth]{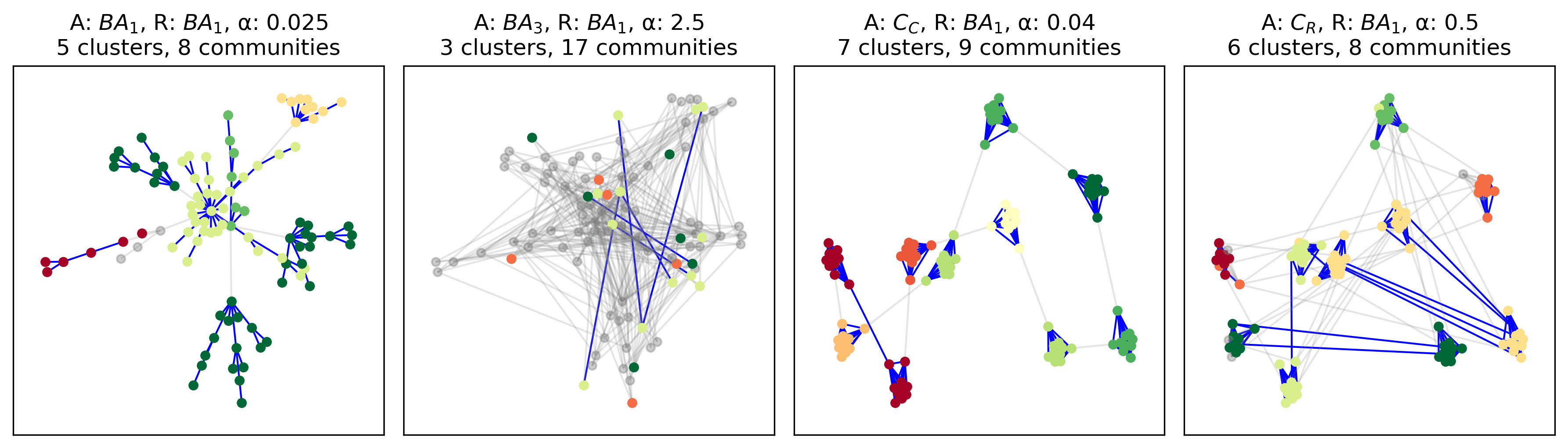}
    \caption{Topological communities in the 4 example networks in Figure \ref{fig:4_example_nets}, with the model run for $10^5$ time units. 
    Nodes are coloured by cluster, corresponding with Figure \ref{fig:4_example_results}. Intra-cluster links are blue, inter-cluster links are grey. The number of clusters and topological communities for each network are given in the title.}
\label{fig:4_example_comms}
\end{figure}

The approach to identify clustering is network topology-agnostic. Thus it is ambiguous whether opinions form a cluster because they have influenced each other directly, or due to coincidence and/or second-order (or greater) effects. Section \ref{sec:setup} briefly defined the concept of a topological community as the connected components after unclustered nodes and inter-cluster links are removed. We illustrate this in Figure \ref{fig:4_example_comms}, for the same example networks and normalised model results used in Figures \ref{fig:4_example_nets}--\ref{fig:4_example_results}. Nodes are coloured by cluster (or grey for unclustered) following Figure \ref{fig:4_example_results}. Intra-cluster links are blue, and inter(non)-cluster links are grey. 
For ${\cal A} \in BA_1$ on the left, 5 clusters are detected in the results, but some of these are comprised of distinct topological networks with grey links, leading to 8 communities (with one red community consisting of a single node). Due to the structural cliques intrinsic to $C_C$ and $C_R$, 
we would intuitively expect 10 topological communities, but fewer clusters are detected (7 and 6, respectively) due to common opinions between different communities. Some of these communities with similar opinions are connected, while others are not (for instance, there are orange nodes amongst the red cluster in the ${\cal A} \in C_R$ results), leading to a count of 9 and 8 topological communities, respectively. Finally, for ${\cal A} \in BA_3$, most nodes are unclustered, and of the nodes that are clustered many of them are not topologically connected; since this network is relatively homogeneous, it is usually coincidence that nodes end up with similar opinions, leading to 17 topological communities (of which many are single nodes) within the 3 clusters.

\FloatBarrier
\subsection{Results}

Following the final opinion distribution classification described in \ref{sec:classification}, we perform a detailed analysis of 4 different network types for ${\cal A} \in \{BA_1$, $BA_3$, $C_C$ and $C_R\}$, and ${\cal R} \in BA_1$. Reiterating, each value of $\alpha$ contains $10^3$ different instances for ${\cal A} \in \{BA_1,BA_3,C_R\}$, to account for the variability of ${\cal A}$, ${\cal R}$, and the initial conditions. The case ${\cal A} = C_C$ contains 100 different instances per $\alpha$-value due to lack of variability in the connected caveman graph. There are two kinds of figures: stacked histograms, showing the distribution of results for each $\alpha$, and scatter plots, showing the average results as a function of $\alpha$ with vertical lines representing standard deviation, and each result (consistently) colour coded by the most prevalent paradigm.

\begin{figure}[!ht]
    \centering
\includegraphics[width=1\textwidth]{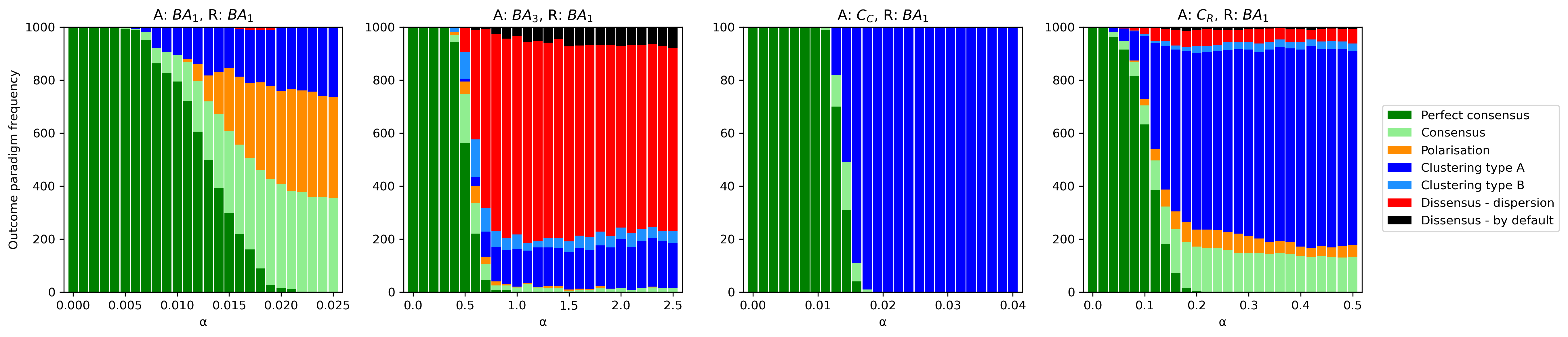}
    \caption{Stacked histograms showing the frequency of classification within each paradigm for the four network pairs as a function of $\alpha$, with colours corresponding to paradigm. Dark green: perfect consensus, light green: consensus, yellow: polarisation, dark/light blue: clustering, red/black: dissensus.}
\label{fig:4_paradigms}
\end{figure}

The frequency of each paradigm result for the 4 pairs is presented in Figure \ref{fig:4_paradigms}. For ${\cal A} \in BA_1$, perfect consensus (dark green) gives way to other paradigms at a very low value of $\alpha$, consistent with results in Section \ref{sec:different}. This transition results in a spread of consensus (light green), polarisation (yellow) and clustering paradigms (blue). This spread arises due to the relatively high amounts of edge betweenness centrality, causing model sensitivity to any particular ${\cal A} \in BA_1$ instance. Although we assign a single paradigm in the results (based on the order presented in Section \ref{sec:classification}), there is often clustering within the polarised groups where the paradigm is polarisation, and consensus sometimes results from a cluster with more than 50\% of nodes on one side of a polarising gap. Assigning sub-classifications would highlight such nuances, but add to the complexity of presenting the results.
For ${\cal A} \in BA_3$, the transition from perfect consensus leads to a mix of predominantly dissensus (dispersion) and a lower rate of clustering, 
due to $\Delta_{i,i+1}$ values rising above the clustering threshold. 
For ${\cal A} \in C_C$, the transition from global consensus quickly sees 100\% occurrence of clustering emerging due to the sharply defined network structure, with minor instances of consensus appearing on the boundary. Finally, the ${\cal A} \in C_R$ histogram shows a mix of clustering, consensus, dissensus (dispersion) and polarisation after the transition, since the random variability in $C_R$ leads to weaker clustering than for $C_C$, though clustering is still the prevalent outcome.

\begin{figure}[!ht]
    \centering
\includegraphics[width=1\textwidth]{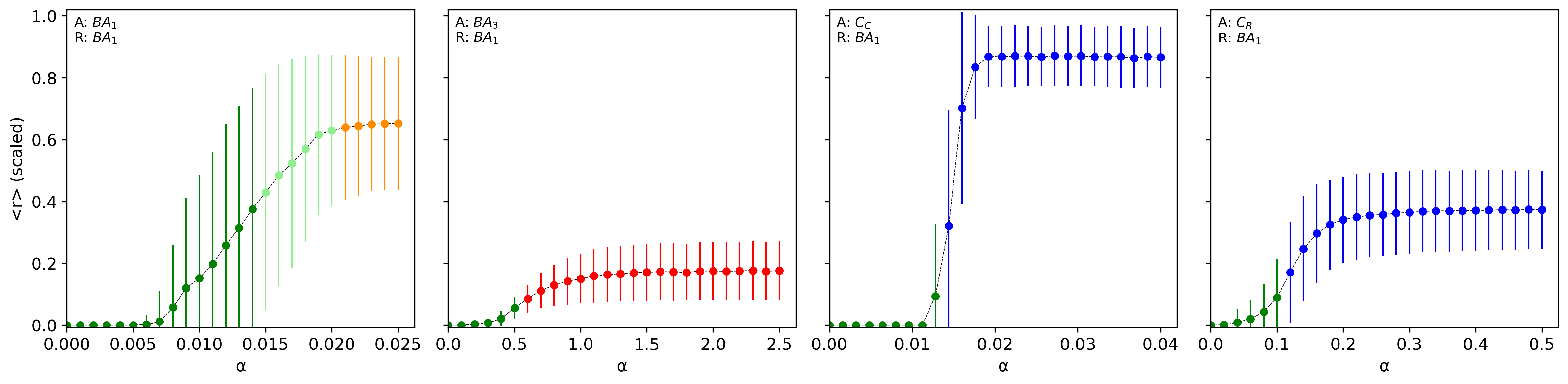}
    \caption{Plots of average order parameter (after scaling and excluding outliers, with error bars representing the standard deviation) against $\alpha$ for the four network pairs, with colours corresponding to modal paradigm as per Figure \ref{fig:4_paradigms}.}
\label{fig:4_metrics_scaled}
\end{figure}

The order parameter results, after scaling and excluding outliers, are shown in Figure \ref{fig:4_metrics_scaled}. Though the equivalent plots in Figure \ref{fig:ARdiff} clearly highlight the transition away from perfect consensus, they are dominated by the macroscopic scale of the opinion differences. Figure \ref{fig:4_metrics_scaled} better enables quantitative comparison of behaviour and correlation with the modal opinion paradigm as a function of $\alpha$. 
Notably, ${\cal A} \in BA_3$ sees a dispersion of opinions and a low scaled order parameter, whilst ${\cal A} \in C_C$ has very strong clustering and very high scaled order parameter. These stark differences arise largely due to the topology of each ${\cal A}$ graph type. Specifically, since $BA_3$ is approximately homogeneous, increasing controversialness drives opinions apart to an approximately continuous spectrum, resulting in dispersion. Likewise, the barely-connected cliques of $C_C$ naturally result in clustering, where the inter-clique ${\cal R}$ links drive entire communities apart. Of the remaining two pairs, ${\cal A} \in BA_1$ typically has higher order parameter than ${\cal A} \in C_R$; both have clustering, but the greater inter-cluster linkages in the latter case spread the clustered opinions into a more diffuse spectrum, while the individual inter-cluster linkages in the former lead to clear similarities with ${\cal A} \in C_C$ results.


\begin{figure}[!ht]
    \centering
\includegraphics[width=1\textwidth]{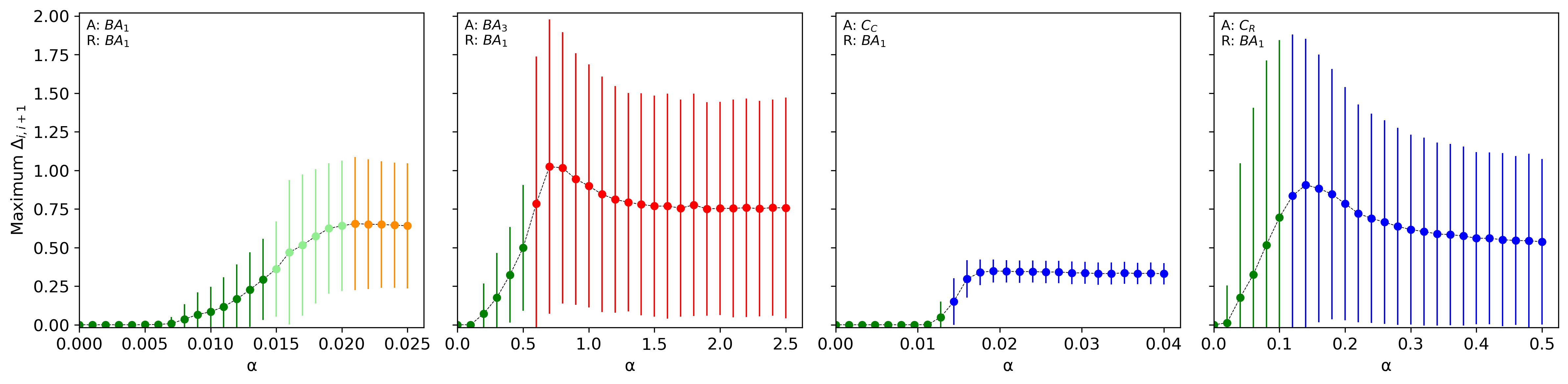}
    \caption{Plots of average largest $\Delta_{i,i+1}$ values (after final opinions were normalised, with error bars representing the standard deviation) against $\alpha$ for the four network pairs, with colours corresponding to modal paradigm as per Figure \ref{fig:4_paradigms}.}
\label{fig:4_delta}
\end{figure}

The average largest $\Delta_{i,i+1}$ values (normalised but including outliers) are shown in Figure \ref{fig:4_delta}. For ${\cal A} \in BA_1$, the maximum $\Delta_{i,i+1}$ values increase with $\alpha$, stabilising just as polarisation becomes the dominant paradigm, averaging at approximately 0.6 (above $Th_{pol}$). In contrast, for ${\cal A} \in BA_3$ and ${\cal A} \in C_R$, the average largest $\Delta_{i,i+1}$ value rises well above $Th_{pol}$, even though polarisation is uncommon (see Figure \ref{fig:4_paradigms}), implying deleting outliers significantly impacts model results. 
Finally, the largest $\Delta_{i,i+1}$ values are relatively low for ${\cal A} = C_C$, since intra-cluster $\Delta_{i,i+1}$ values are small, inter-cluster $\Delta_{i,i+1}$ values are evenly-spaced and below $Th_{pol}$, and outliers are uncommon.

\begin{figure}[!ht]
    \centering
\includegraphics[width=1\textwidth]{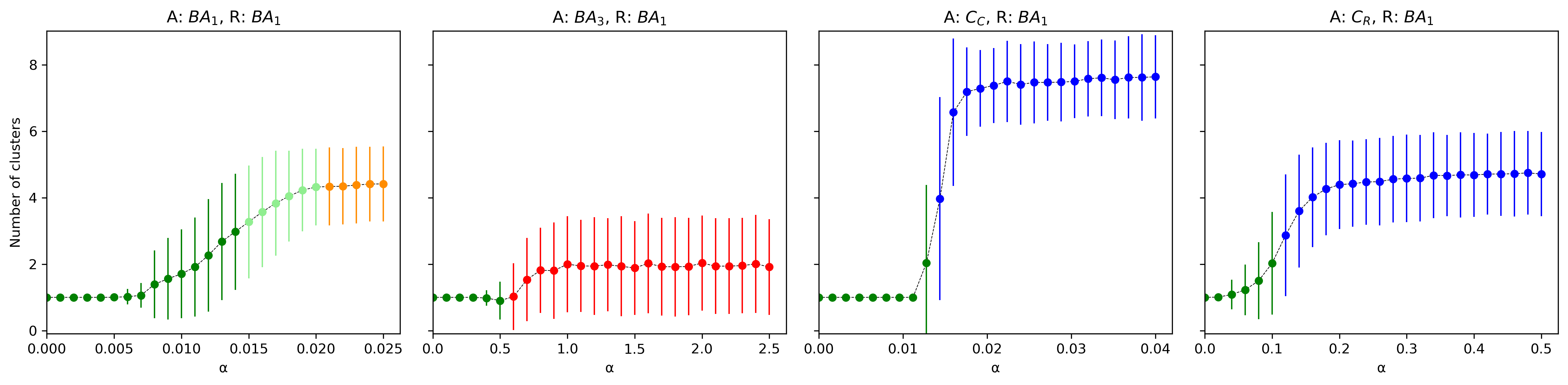}
    \caption{Plots of number of clusters (after outlier deletion and final opinions are normalised, with error bars representing the standard deviation) against $\alpha$ for the four network pairs, with colours corresponding to modal paradigm given in Figure \ref{fig:4_paradigms}.}
\label{fig:4_n_clusters}
\end{figure}

The average cluster counts (based on outlier deletion, and scaling) are shown in Figure \ref{fig:4_n_clusters}. These results follow similar trends to the scaled order parameter in Figure \ref{fig:4_metrics_scaled}. Intuitively, for all graphs, when the paradigm is perfect consensus there is a single cluster. 
For ${\cal A} \in BA_1$ at higher $\alpha$ values, the paradigm is either polarisation, consensus, or clustering, resulting in several clusters. For ${\cal A} \in BA_3$, there are an average of 2 clusters detected within the paradigm of dissensus (dispersion), with these clusters arising from a significant sequence of $\Delta_{i,i+1}$ values rising above the threshold ($Th_{pol}$), rather than being visibly clearly defined clusters (as per the example in Figure \ref{fig:4_example_results}). For ${\cal A} = C_C$, the strong clustering arising from the clique-based topology yields an average of 8 clusters, which is less than the 10 topological clusters because of adjacent cliques sometimes ending up in the same cluster. For ${\cal A} \in C_R$, the number of clusters is lower than for ${\cal A} = C_C$ because of the high inter-cluster linkages spreading the clustered opinions and making them more likely to overlap within the $\Delta_{i,i+1} < 0.05$ threshold to form a single cluster.

\begin{figure}[!ht]
    \centering
\includegraphics[width=1\textwidth]{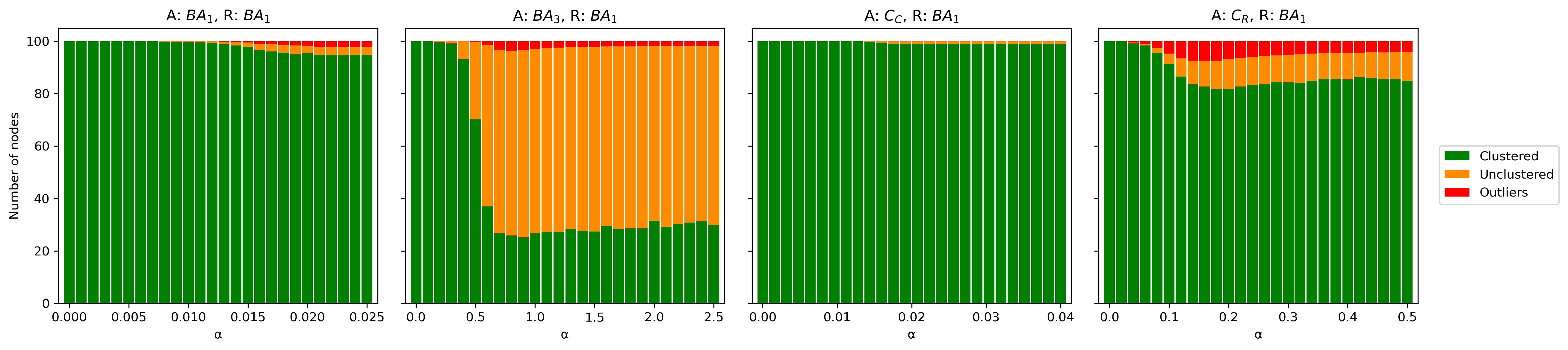}
    \caption{Stacked bar charts showing the average number of clustered (green), unclustered (yellow) and outlier (red) nodes for the four network pairs as a function of $\alpha$.}
\label{fig:4_clustered_nodes}
\end{figure}

To better understand the cluster compositions counted in Figure \ref{fig:4_n_clusters}, the average number of nodes identified as clustered, unclustered or outliers for each pair of graphs and each $\alpha$ value is shown in Figure \ref{fig:4_clustered_nodes}. For ${\cal A} \in BA_1$ and ${\cal A} = C_C$, the vast majority of final opinions fall into clusters, owing to their topologies possessing minimal inter-cluster links in ${\cal A}$. The number of clustered nodes is comparatively lower for ${\cal A} \in C_R$ since some opinions become linked to multiple clusters and are pulled equally between them (thus unclustered), or else weakly bound to a cluster and strongly repelled by multiple links in ${\cal R}$ --- becoming outliers. In contrast, the majority of nodes are unclustered for ${\cal A} \in BA_3$ due to the regular topology.

\begin{figure}[!ht]
    \centering
\includegraphics[width=1\textwidth]{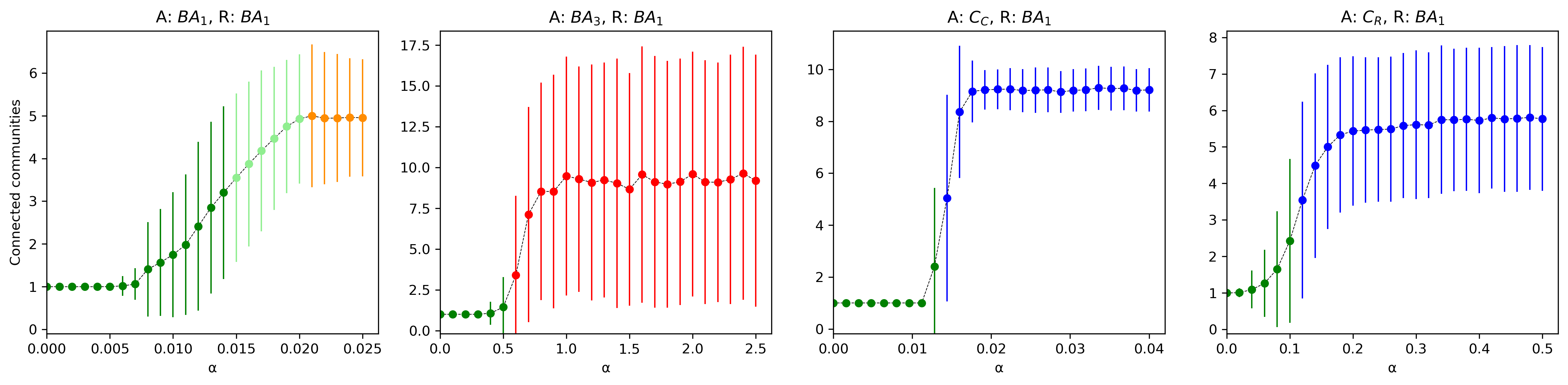}
    \caption{Plots of number of topological communities (connected components after unclustered nodes and inter-cluster links are removed) against $\alpha$ for the four network pairs, with colours corresponding to modal paradigm as per Figure \ref{fig:4_paradigms}.}
\label{fig:4_n_communities}
\end{figure}

Investigating the interrelationship between ${\cal A}$ and the formation of opinion clusters, results of the number of topological communities (defined in Section \ref{sec:setup} as connected components after unclustered nodes and inter-cluster links are removed) are presented in Figure \ref{fig:4_n_communities}. 
For ${\cal A} \in BA_1$ and ${\cal A} = C_C$, the number of communities is slightly higher than the number of clusters, since 2 or more connected components on opposite sides of the network occasionally form similar opinions; in the latter case, the number of topological communities gets quite close to the number of structural cliques (10). The number of communities is slightly higher than the number of clusters for ${\cal A} \in C_R$ as well, though the randomness in link-rewiring causes a higher standard deviation and the resulting higher interconnectedness of corresponding cliques prevents the number of communities from reaching the number of cliques. In contrast to these 3 attraction graphs, which often fall within the clustering paradigm, the number of communities ($\approx 10$) is significantly higher than the number of clusters ($\approx 2$) for ${\cal A} \in BA_3$. Additionally, the standard deviation for this case is relatively large, resulting in 15 communities within one standard deviation. When paired with the results in Figure \ref{fig:4_n_clusters} and Figure \ref{fig:4_clustered_nodes}, this indicates that it is common for 2 clusters to be identified each with around 10 nodes, but of these 20 nodes there may be 10 connected components, i.e. only 10 pairs of 2 linked nodes with similar opinions. This strengthens the need to perform the detailed analysis performed in this work to expose nuanced cases of clustering coincidentally arising potentially due to approximately homogeneous topologies.   


\begin{figure}[!ht]
    \centering
\includegraphics[width=0.45\linewidth]{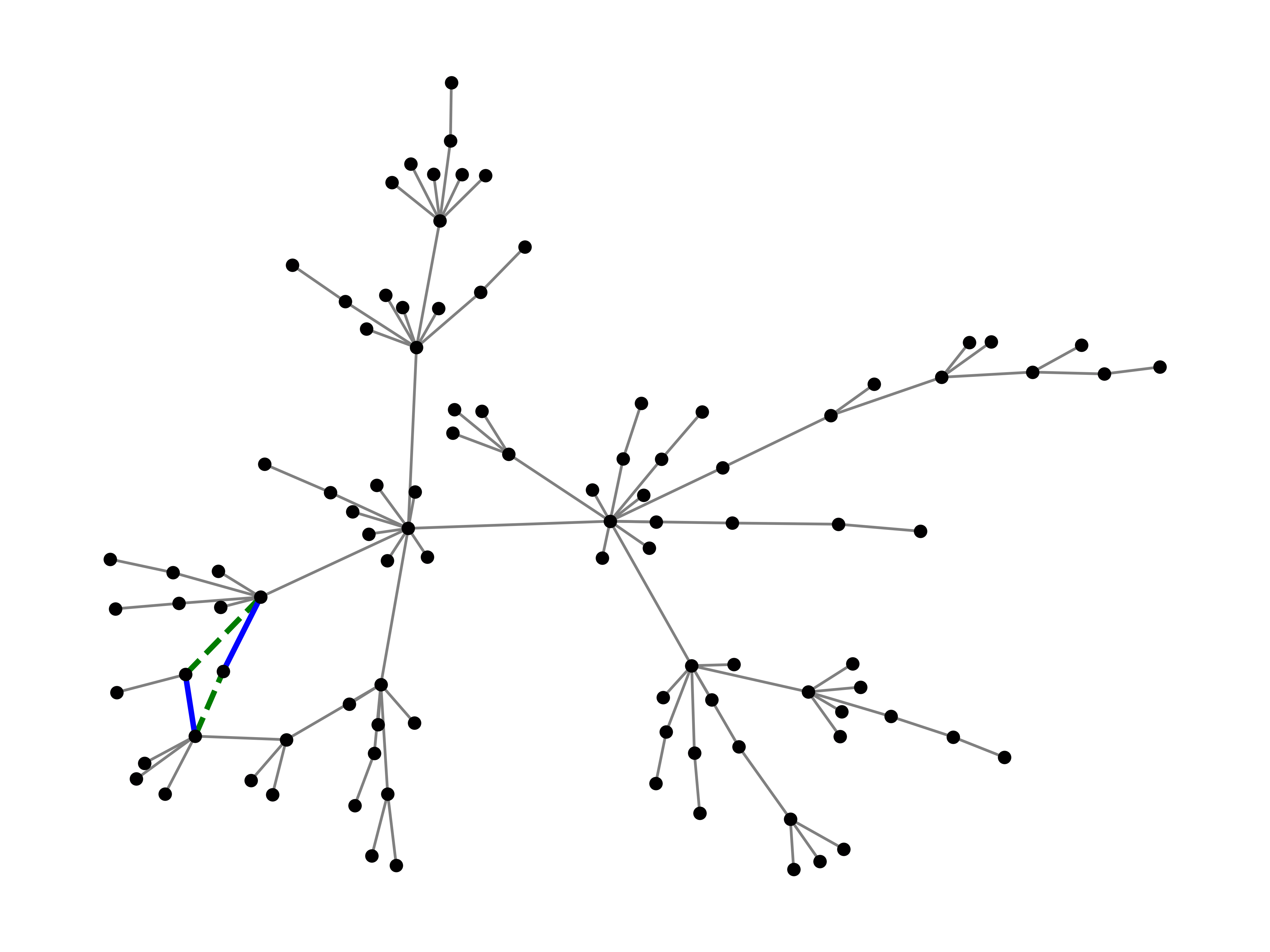}
\includegraphics[width=0.45\linewidth]{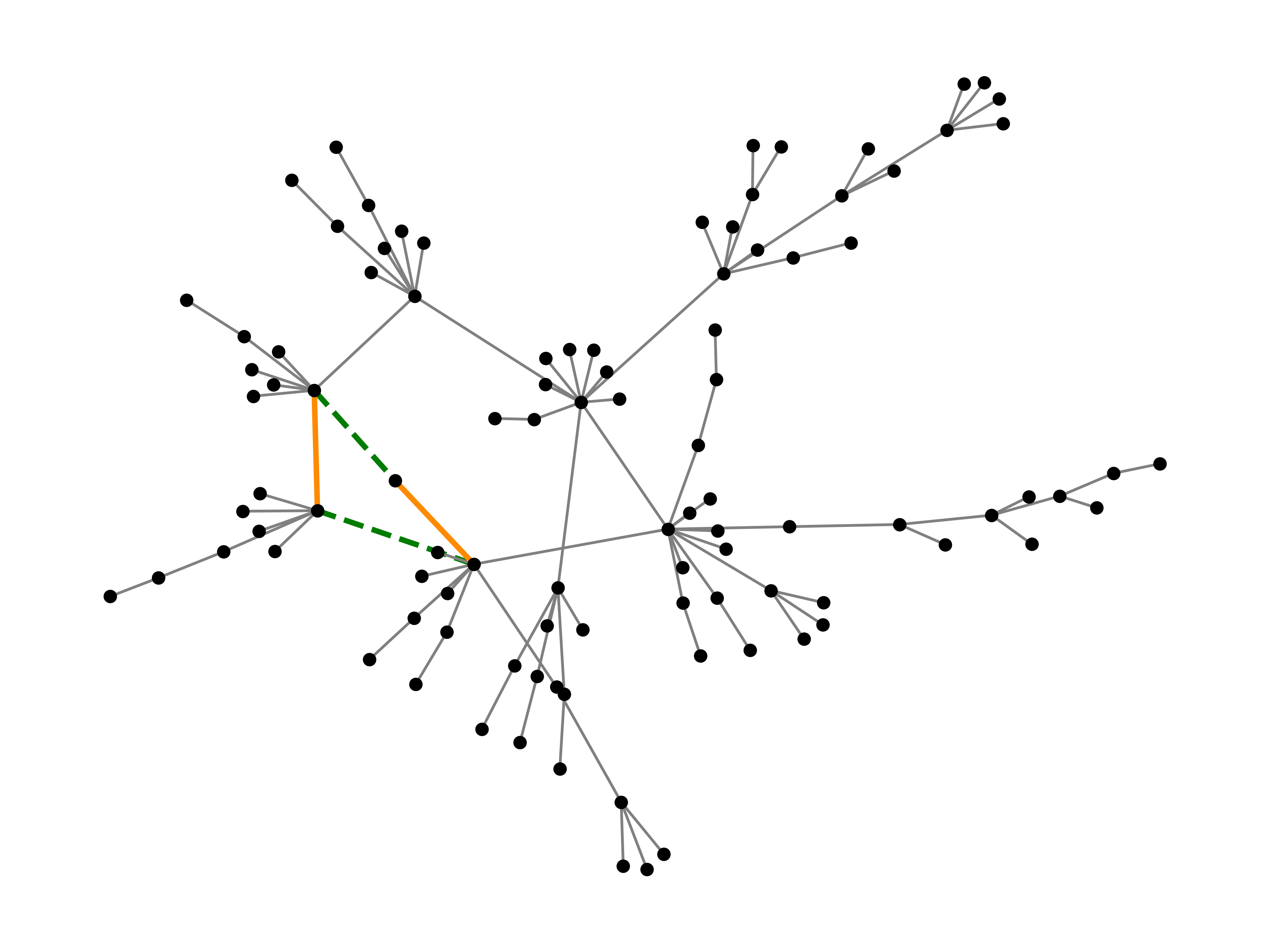}
    \caption{Two examples of the connected double edge swap algorithm on a given network ${\cal A}$. Edges that changed are coloured. The paradigm before making any swaps was perfect consensus (green dashed lines). The swapped edges on the left (blue) caused the paradigm to shift to clustering (type A) and on the right, changing a different pair of edges (yellow) resulted in polarisation. Here, the original ${\cal A}$, ${\cal R}$ networks were $BA_1$, with identical initial conditions and $\alpha = 0.14$. One edge pair within both ${\cal A}$ and ${\cal R}$ was swapped whilst preserving degree and keeping the networks connected. }
\label{fig:edge_swap_network}
\end{figure}

To examine how network topology impacts the paradigm, a minimal perturbation was applied to networks ${\cal A}$ and ${\cal R}$ by using an edge swap algorithm before rerunning the model and checking for different outcomes.
Two examples of the algorithm on network ${\cal A}$ for $\alpha = 0.14$ are shown in Figure \ref{fig:edge_swap_network}.
The base networks ${\cal A}$ and ${\cal R}$ (both $BA_1$) for given initial conditions $x_0$ lead to perfect consensus (green dashed lines).
After swapping ${\cal A}$'s green edges with blue (left) and yellow (right) edges, and applying similar swaps to ${\cal R}$ whilst keeping all other values consistent, the paradigm changes to clustering and polarisation respectively.
These findings show that even small changes to network topology can have significant impacts.


\begin{figure}[!ht]
    \centering
\includegraphics[height=3.3cm]{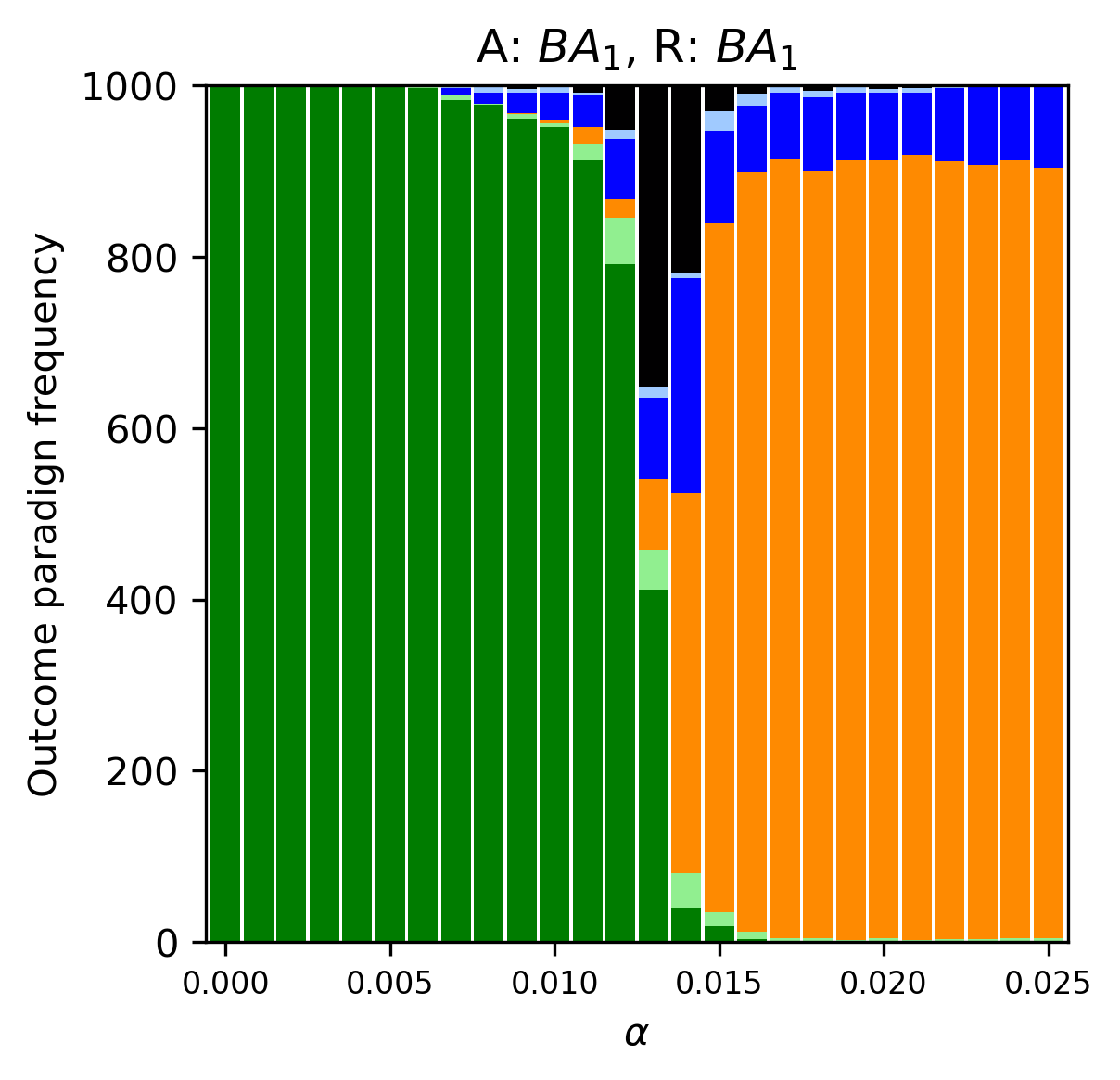}
\includegraphics[height=3.3cm]{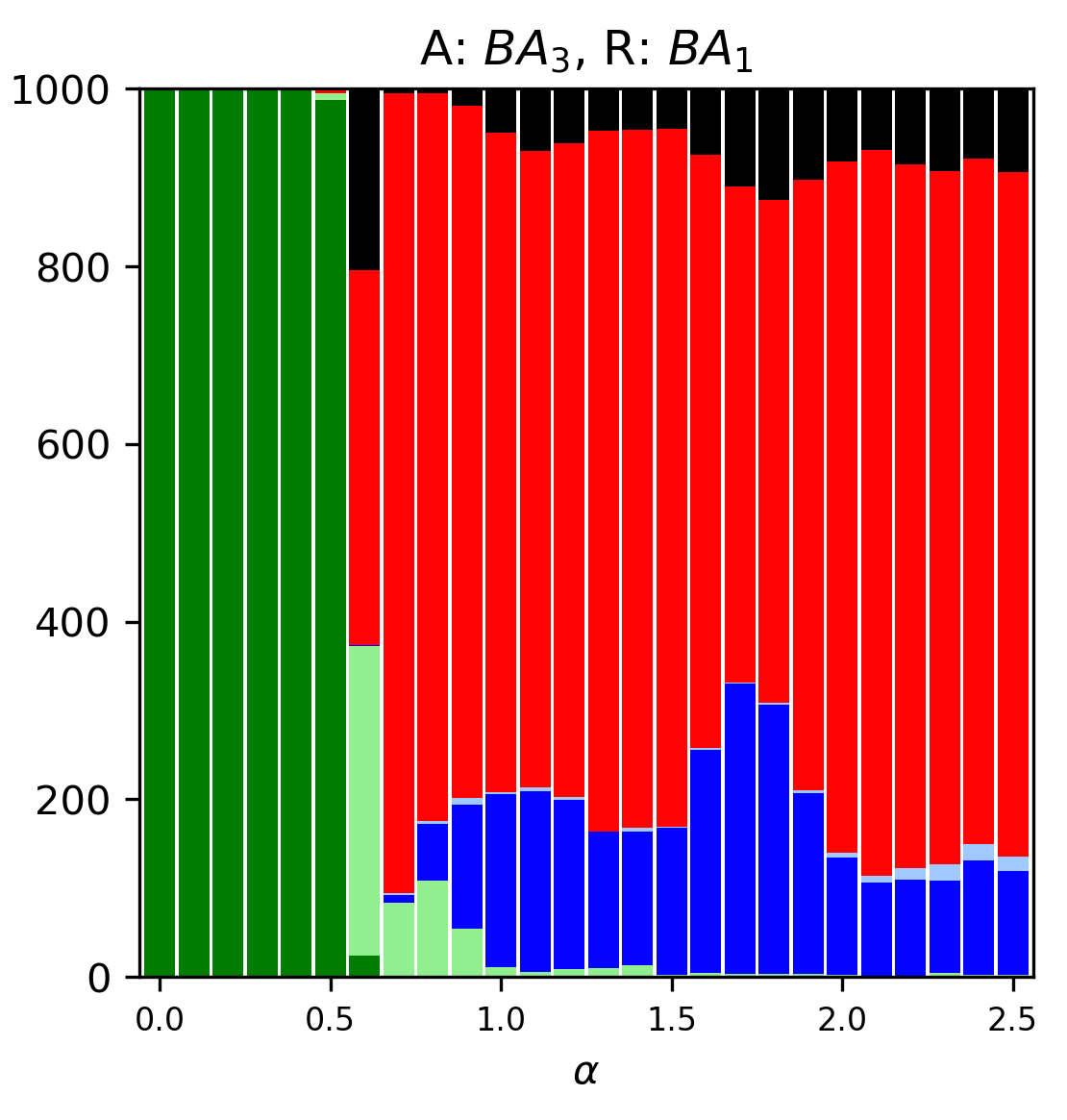}  
\includegraphics[height=3.3cm]{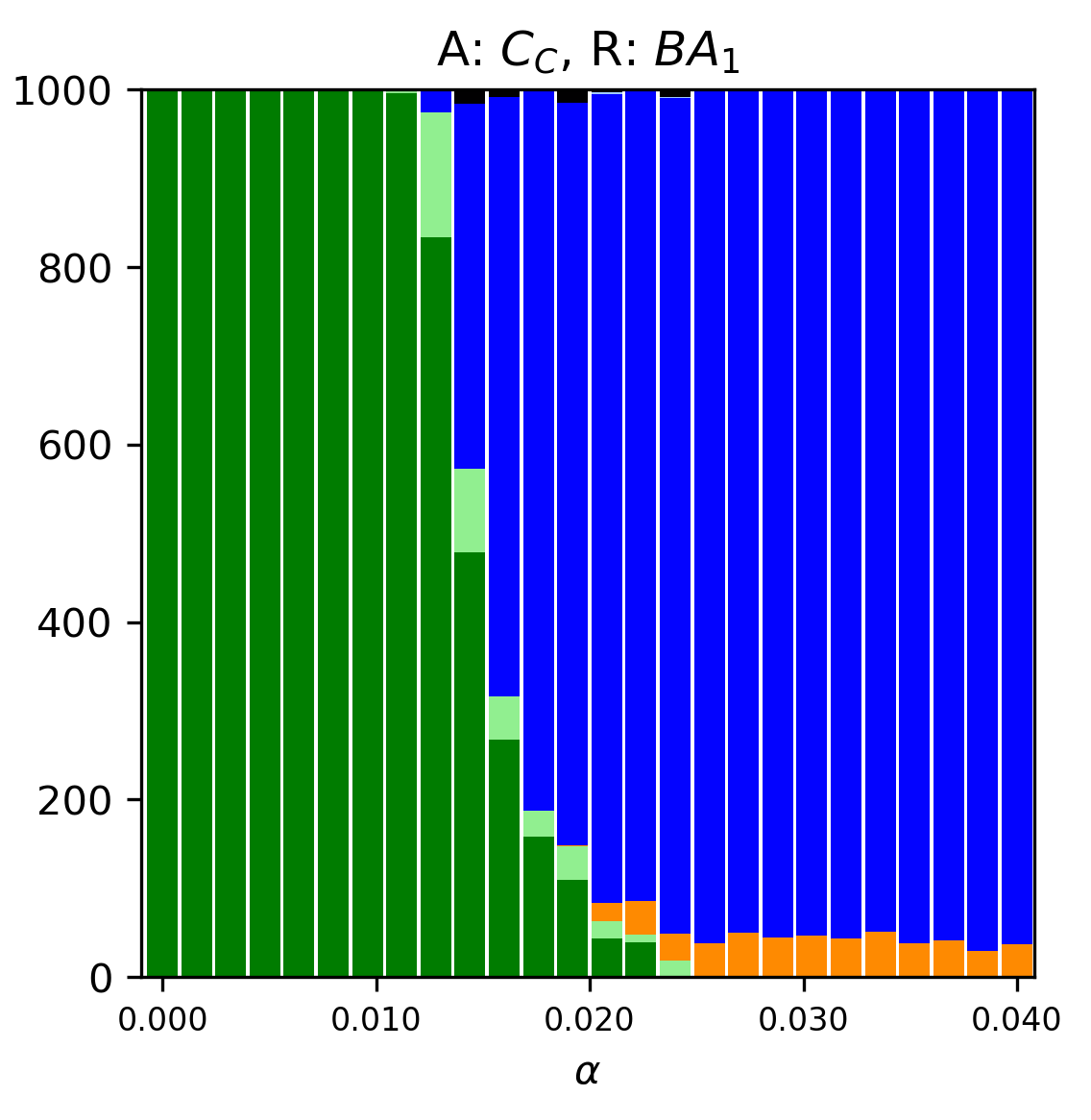}
\includegraphics[height=3.3cm]{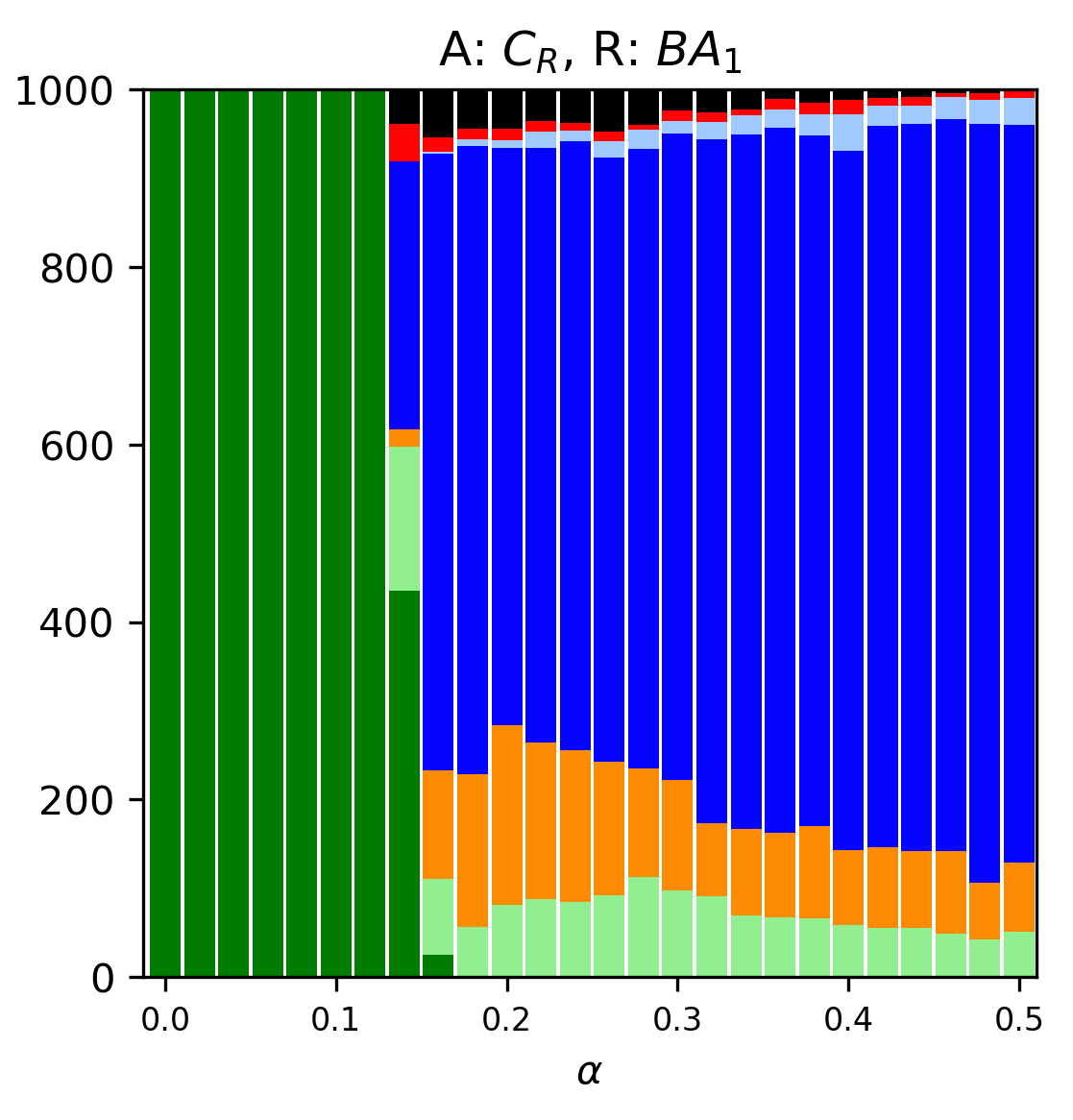}
\includegraphics[height=3.3cm]{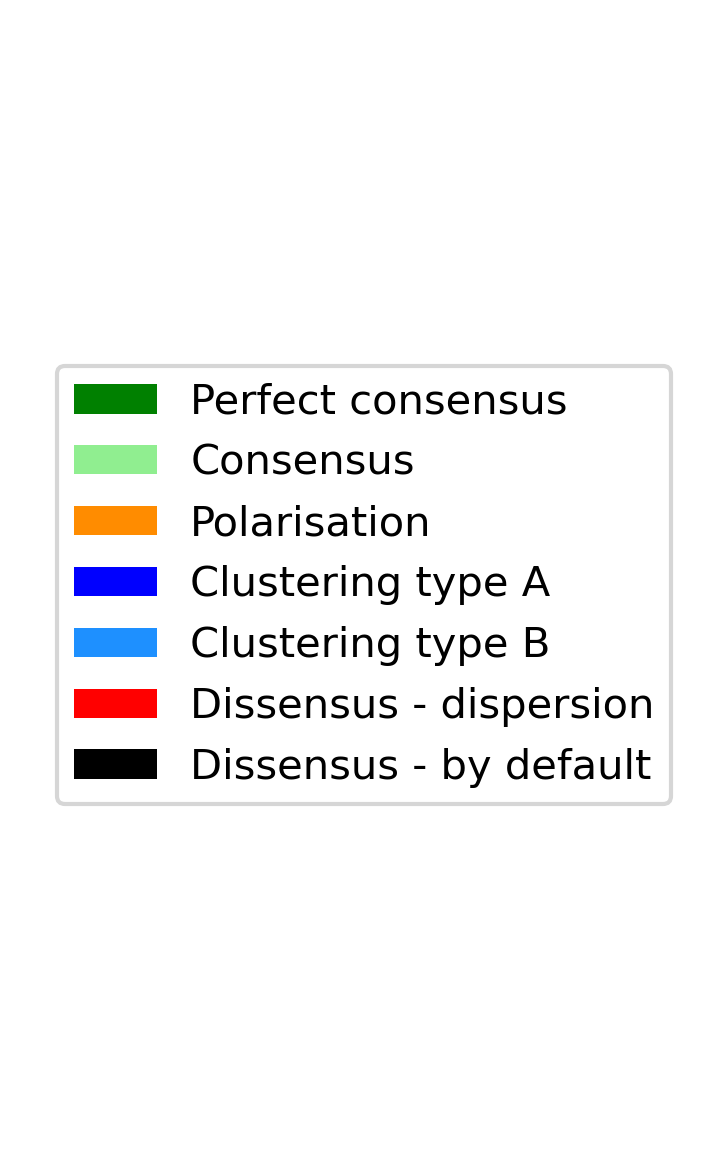}
    \caption{Stacked histograms depicting the impact of varying network topology on the resulting paradigm for given attraction and repulsion networks. For each plot, the base ${\cal A}$, ${\cal R}$ networks were fixed, initial conditions remained the same, and one edge pair within both ${\cal A}$ and ${\cal R}$ was swapped. This was iterated 1000 times for each $\alpha$ value.}
\label{fig:edge_swap_paradigm}
\end{figure}

A more detailed exploration of this behaviour is shown in Figure \ref{fig:edge_swap_paradigm}, which sweeps across 26 $\alpha$ values for each instance.
Where Figure \ref{fig:4_paradigms} bootstrapped 1000 randomised networks per $\alpha$ value, each plot in Figure \ref{fig:edge_swap_paradigm} has a fixed ${\cal A}$ and ${\cal R}$ network, and one double edge swap is applied to both networks per iteration.
It is important to note that in all of these plots, the initial conditions $x_0$ were fixed, so all changes observed for a particular $\alpha$ value are purely due to changes in the network topology.
This form of perturbation can be interpreted in the context of an online social network by individuals `following' (`friending') or `unfollowing' (`unfriending') people. 
To preserve each individual's total number of connections (the degree distribution), the algorithm applies a swap rather than adding or removing edges.

For the leftmost panel, ${\cal A} \in BA_1$, all paradigms except dissensus (dispersion) are observed across the range of $\alpha$ values.
For $\alpha < 0.13$, the dominant paradigm is consensus, whereas for $\alpha > 0.15$, most edge swaps result in polarisation.
The intermediate values of $\alpha$ display a variety of paradigms, potentially due to the tree-like structure of the $BA_1$ graph strongly driving the dynamics of groups of nodes together or apart depending on which edges are swapped. 
The impacts of edge swapping on ${\cal A} \in BA_3$ are less pronounced; likely because there are more edges in the graph's inter- and intra-cluster groups.
Thus, slight topological perturbations are unlikely to impact information flows across the network (and in turn, the paradigm).
As for the randomised graphs in Figure \ref{fig:4_paradigms}, the most observed paradigm is dissensus by dispersion.
In addition, there are fewer emergent behaviours observed for ${\cal A} \in BA_3$ at $\alpha \simeq 0.6$ where the dominant paradigm shifts from consensus to dispersion.
This is likely due to the topological variety arising from one swapped edge pair being less than that of generating a new $BA_3$ random graph.
Finally, the behaviour for ${\cal A} \in C_C$ and ${\cal A} \in C_R$ (two right panels) is close to what is observed in Figure \ref{fig:4_paradigms}, but the proportions of the paradigms display a slight shift towards the dissensus end of the spectrum, i.e. consensus instead of perfect consensus, or polarisation instead of consensus.
This can be explained by the fact that the majority of edges in the cavemen graphs are in cliques; random changes to inter-cluster edges alter the regular structure of the caveman network with corresponding dynamical consequences.


\section{Conclusions}
\label{sec:conclusion}
Models of opinion dynamics have been actively studied since the 1940s, and their relevance has only increased with the advent of the modern internet and social media.
In a contemporary online setting, theoretically, any individual may become an opinion leader and manipulate opinions of large populations. 
The utility of quantitative models in exploring opinion dynamics has only become more important as online influencers begin to have impacts on decision making in topics such as  health and politics \cite{Pagan21}.
Here, we focus on trolls, who destabilise and divide communities through the spread of toxic, false, or controversial narratives.

To better understand the impact of trolling, this paper offers a novel modelling framework that captures the effect of controversy on opinions whilst tracking community interactions through attraction and repulsion networks.
Our model indicates that strong clustering in the underlying network structure drives clustering of opinions, and that controversy is particularly divisive when it arises from socially distant, rather than socially close, peers.
A key finding is that as controversialness  increases, it is possible to observe a critical threshold where consensus starts to break down.
More broadly, the modelling results indicate that an interplay between initial conditions, network topology, and the level of controversy drive emergent behaviours. 
We anticipate that future work will further untangle these relationships and explore the model's applicability to real-world data, providing further insight into the factors which allow controversial narratives to divide communities.

\section*{Acknowledgements}
We are grateful for the financial support provided by Defence Innovation Partnership and Defence SA through the Collaborative Research Fund project ``Mathematical modelling of a complex and contested conflict environment: influence of logistics and resource transfer''. JCMC is the recipient of an Australian Research Council Early Career Industrial Fellowship (project number IE240100140) funded by the Australian Government.

\appendix
\section{Graph-Laplacian analysis} \label{app:lap}
Using the identity 
\begin{eqnarray}
   \sum^N_{j=1}{\cal A}_{ij} (x_i-x_j) &=&  \sum^N_{j=1}\left({\cal D}^{{\cal A}}_{ij} -{\cal A}_{ij} \right)x_j \equiv \sum^N_{j=1}{\cal L}^{{\cal A}}_{ij} x_j \label{laplacian}
\end{eqnarray}
where ${\cal D}^{{\cal A}}$ is the \textit{degree}-matrix and ${\cal L}^{{\cal A}}$ is the \textit{Laplacian} corresponding to ${\cal A}$, Equation (\ref{same2}) becomes 
\begin{equation}
       \dot{x}_i = - \frac{(1-\alpha)}{N}\sum^N_{j=1}{\cal L}^{{\cal A}}_{ij} x_j, \;\; i \in \{1,\dots,N\}, \;\; \alpha \in \mathbb{R}_+.\label{same7}
\end{equation}
with corresponding eigenvalue spectra given by Eq.(\ref{spec1}), and orthonormal eigenvectors --- labelled $\nu^{(r)}_i$ --- possessing the properties
\begin{eqnarray}
    \sum^N_{j=1}{\cal L}^{{\cal A}}_{ij} \nu^{(r)}_j = \lambda^{\cal A}_r \nu^{(r)}_i, \;\;\; \sum^N_{j=1} \nu^{(r)}_j \nu^{(s)}_j = \delta_{rs} \label{e-iden}
\end{eqnarray}
where the zeroth eigenvector elements are always of the form,  
\begin{equation}
    \nu^{(0)}_i = \frac{1}{\sqrt{N}},\;\; i \in \{1, \dots, N\}
\end{equation}
Expanding the node variables $x_i$ as the following sum of normal modes $y_r$ via
\begin{eqnarray}
    x_i = \sum^{N-1}_{r=0} \nu^{(r)}_i y_r, \;\;\; y_r = \sum^{N}_{i=1} \nu^{(r)}_i x_i \label{node-mode}
\end{eqnarray}
and applying the identities in Equation (\ref{e-iden}), Equation (\ref{same7}) collapses to
\begin{eqnarray}
    \dot{y}_r = - \frac{(1-\alpha)\lambda^{\cal A}_r}{N} y_r \;\;\; \Rightarrow \;\;\; y_r(t) = y_r(0) \cdot e^{-\frac{(1-\alpha)\lambda^{\cal A}_r}{N} t} \label{stable}
\end{eqnarray}
Equation (\ref{stable}) shows that solutions to Equation (\ref{same}) are exponentially stable if $\alpha < 1$. Thus the solutions to the original node variables $x_i$ are
\begin{eqnarray}
    x_i(t) = \sum^{N-1}_{r=0} \sum^N_{j=1} \nu^{(r)}_i\nu^{(r)}_j x_j(0) \cdot e^{-\frac{(1-\alpha)\lambda^{\cal A}_r}{N} t} \label{dyn1}
\end{eqnarray}
which collapses to Eq.(\ref{steadylin}) in the $t\rightarrow \infty$ limit.

\section{Statistical analysis of star graph} \label{app:star}
\begin{figure}[!ht]
    \centering
\includegraphics[width=0.4\textwidth]{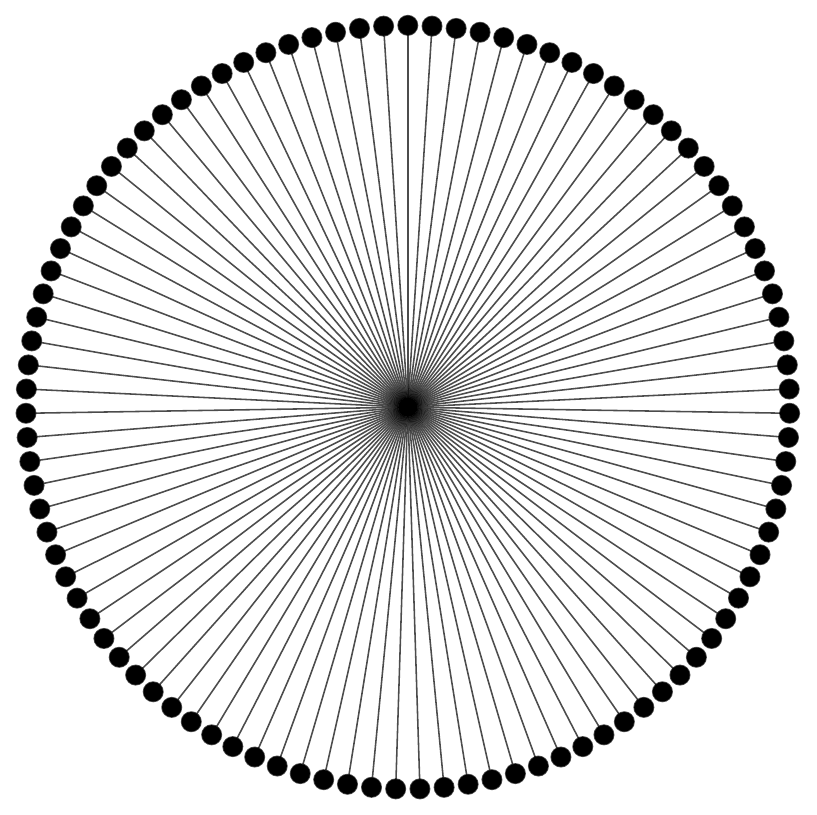}\includegraphics[width=0.6\textwidth]{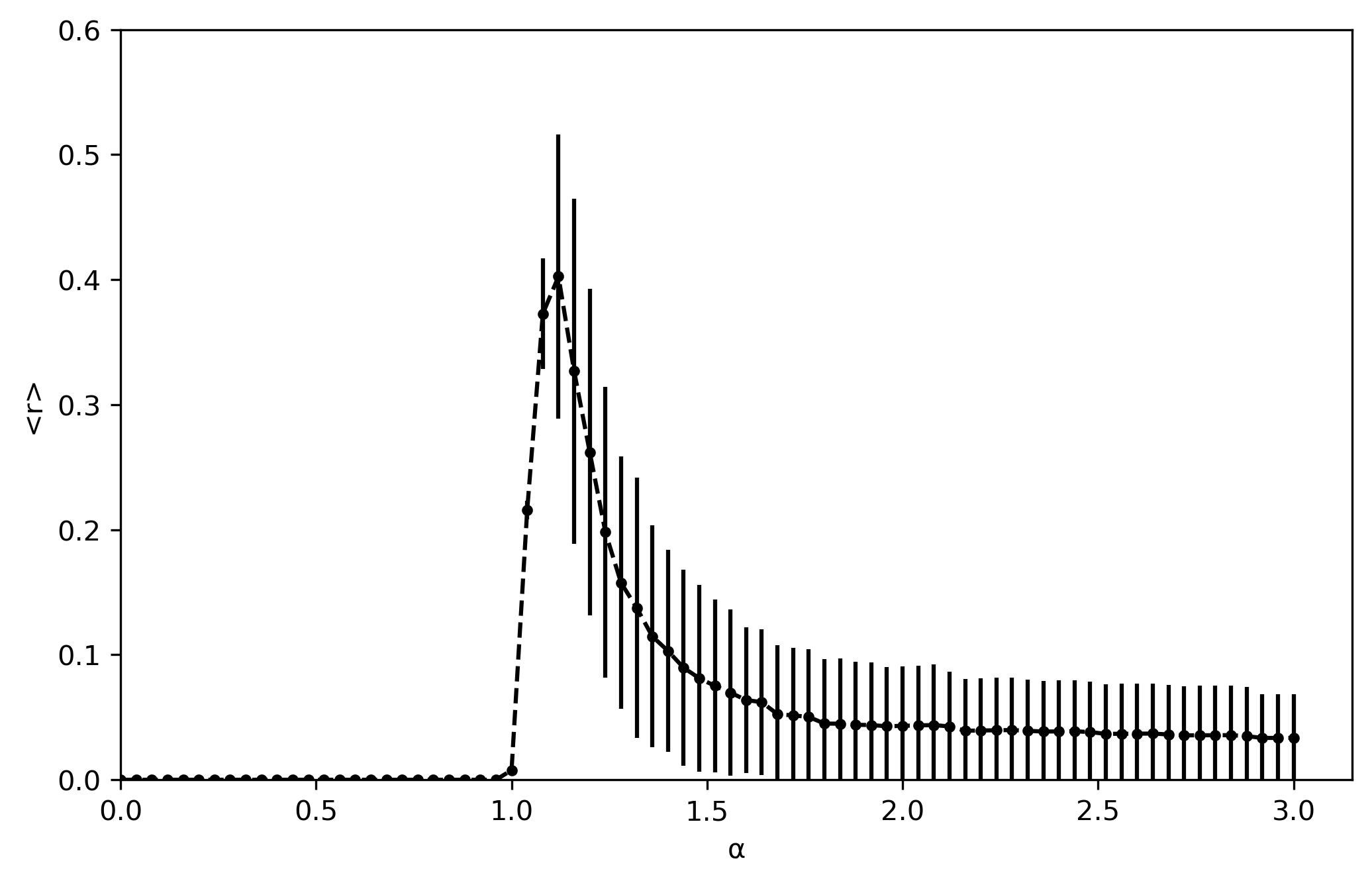}
    \caption{Network structure for the unique star graph (left) and the average results (with error bars representing the standard deviation) of the order parameter as \(\alpha\) increases (right).}
\label{fig:star}
\end{figure} 
Focusing on Figure \ref{fig:star}, the right panel presents the order parameter results, for the particular \textit{star graph} topology which is shown of the left. The star graph is the most extreme instance of preferential attachment displayed in $BA_1$ graphs --- with one node attached to all others, who possess no connections amongst themselves --- and thus deserves its own analysis. The order parameter values $\langle r \rangle$ are obtained by collecting values for Eq.(\ref{order}) over 100 instances of initial conditions. The trajectory gives the average value of $r$ per $\alpha$-value, with the error bars denoting the standard deviation over the initial conditions. Notably, since the attraction and repulsion graphs are identical, the order parameter is zero for $\alpha <1$, as per the analysis in Section \ref{sec:critical}. For $\alpha> 1$, the order parameter rises sharply, due to the appearance of new fixed points deviating from $x_i = x_j$ $\forall \; \{i, j\} \in \{1,\dots,N\}$. In fact, if we label the central node of the star as $x_{centre}$, the form of the potential for the star graph can be expressed as:
\begin{eqnarray}
    V_{star}(\mathbf{x}) = \frac{1}{N}\sum^N_{\genfrac{}{}{0pt}{}{j=1}{\ne centre}}  \left[ \frac{1}{2}(x_{centre}-x_j)^2 - \frac{1}{\alpha} \ln \cosh \alpha (x_{centre} - x_j) \right]. \label{starpot}
\end{eqnarray}
Eq.(\ref{starpot}) contains $N-1$ independent variables in the form $x_{centre} - x_k$, and thus is the summation of $N-1$ copies of Eq.(\ref{N_2}) for $N=2$. For $\alpha>1$, all opinion trajectories (except $x_{centre}$) fall in one of two equilibrium points which are a fixed distance above and below $x_{centre}$. The error bars demonstrate different initial conditions resulting in different proportions of opinions settling in the fixed points above and below $x_{centre}$. As $\alpha$ increases further, the order parameter decreases. This counterintuitive behaviour is be explained by the transient behaviour of opinions, with a greater proportion of trajectories landing on the same fixed point as $\alpha$ increases.


\section{Large controversialness in all-to-all network} \label{sec:theo}
For the all-to-all network, Eq.(\ref{same}) becomes
\begin{eqnarray}
     \dot{x}_i &=& - \frac{1}{N}\sum^N_{j =1}\left[ (x_i-x_j) -  \tanh \alpha (x_i - x_j) \right]\nonumber\\
     &=& -x_i + \bar{x} (0) + \frac{1}{N}\sum^N_{j =1} \tanh \alpha (x_i - x_j)
     \label{allall}
\end{eqnarray}
where we have applied Eqs.(\ref{conserved}) and (\ref{steadylin}) on the term $\sum^N_{j=1} x_j$. Due to the all-to-all network being isomorphic, initial conditions for this case do not effect the final outcome, hence we can impose 
\begin{equation}
    x_i(0) < x_j(0)
\end{equation}
without loss of generality. Assuming $\alpha \gg 1$, Eq.(\ref{allall}) becomes
\begin{eqnarray}
    \dot{x}_i &=&-x_i + \bar{x} (0) + \frac{1}{N}\underbrace{\sum^N_{j =1} \textrm{sgn} (x_i - x_j)}_{N+1-2i}\nonumber\\
    \Rightarrow x_i(t \rightarrow \infty) &=& \bar{x} (0)+ \frac{N+1-2i}{N}\label{allall1}
\end{eqnarray}
Substituting Eq.(\ref{allall1}) into Eq.(\ref{order}) for the order parameter definition reveals
\begin{equation}
    r = \frac{4}{N^4}\sum^N_{i,j = 1} (i-j)^2 = \frac{2}{3}\left(1-\frac{1}{N^2} \right)
\end{equation}
which is Eq.(\ref{theor}) of the main text. 


\end{document}